\documentclass[reprint,aps,prl,showpacs,floatfix,superscriptaddress,citeautoscript,a4paper,longbibliography]{revtex4-1}

\pdfoutput=1

\usepackage{amsmath, amsfonts, amssymb, amsbsy}
\usepackage[margin=22mm]{geometry}
\usepackage{tipa}
\usepackage{graphicx}
\usepackage{xspace}

\usepackage[caption=false]{subfig}

\makeatletter
\def\tagform@#1{\maketag@@@{\ignorespaces#1\unskip\@@italiccorr}}
\makeatother

\usepackage{hyperref}
\hypersetup{
        pdftitle={Site-dependent charge transfer at the Pt(111)-ZnPc interface and the effect of iodine},
        pdfdisplaydoctitle=true,
        pdfsubject = {Article},
        pdfauthor= {BMW},
        pdfkeywords= {},
        colorlinks=true,
        linkcolor=blue,
        citecolor=blue,
        urlcolor=blue,
        unicode=false
}

\begin{document}
\title{Site-dependent charge transfer at the Pt(111)-ZnPc interface and the effect of iodine}

\author{S.~Ahmadi}
\email{sareha@kth.se}
\affiliation{KTH Royal Institute of Technology, ICT MNF Materials Physics, Electrum 229, 164 40 Kista, Sweden}
\author{B.~Agnarsson}
\affiliation{Department of Applied Physics, Biological Physics, Chalmers University of Technology, Fysikgr\"{a}nd 3, 
412 96 G\"{o}teborg, Sweden}
\author{I.~Bidermane}
\affiliation{Department of Physics and Astronomy, Uppsala University, Box 516, 751 20 Uppsala, Sweden}
\author{B.~M. Wojek}
\affiliation{KTH Royal Institute of Technology, ICT MNF Materials Physics, Electrum 229, 164 40 Kista, Sweden}
\author{Q. No\"{e}l}
\affiliation{KTH Royal Institute of Technology, ICT MNF Materials Physics, Electrum 229, 164 40 Kista, Sweden}
\author{C.~H. Sun}
\affiliation{Australian Institute for Bioengineering and Nanotechnology, The University of Queensland, QLD 
4072 Brisbane, Australia}
\author{M.~G\"{o}thelid}
\email{gothelid@kth.se}
\affiliation{KTH Royal Institute of Technology, ICT MNF Materials Physics, Electrum 229, 164 40 Kista, Sweden}

\date{\today}

\begin{abstract}
The electronic structure of ZnPc, from sub-monolayers to thick films, on bare and iodated
Pt(111) is studied by means of X-ray photoelectron spectroscopy (XPS), X-ray absorption
spectroscopy (XAS) and scanning tunneling microscopy (STM). Our results suggest that at
low coverage ZnPc lies almost parallel to the Pt(111) substrate, in a non-planar configuration
induced by Zn-Pt attraction, leading to an inhomogeneous charge distribution within the
molecule and charge transfer to the molecule. ZnPc does not form a complete monolayer on
the Pt surface, due to a surface-mediated intermolecular repulsion. At higher coverage ZnPc
adopts a tilted geometry, due to a reduced molecule-substrate interaction. Our photoemission
results illustrate that ZnPc is practically decoupled from Pt, already from the second layer.
Pre-deposition of iodine on Pt hinders the Zn-Pt attraction, leading to a non-distorted
first layer ZnPc in contact with Pt(111)-I $\left(\sqrt{3}\times\sqrt{3}\right)$ or Pt(111)-I 
$\left(\sqrt{7}\times\sqrt{7}\right)$, and a more homogeneous charge distribution and charge transfer at the 
interface. On increased ZnPc thickness iodine is dissolved in the organic film where it acts as an electron acceptor 
dopant.
\end{abstract}


\maketitle

\section{Introduction}
\label{sec:intro}
During recent years, studies on the growth of phthalocyanine (Pc) films on different surfaces,
especially metallic substrates have become a major field for the surface science and organic
electronic communities. These molecules can create self-assembled layers, which make the
thin-film processing very convenient\cite{Betti-JPhysChemC-2010, Kroeger-NewJPhys-2010, 
Nilson-JPhysChemC-2010, Aalund-SurfSci-2007, Ahmadi-JChemPhys-2012, Stadler-NatPhys-2009, 
Stadler-PhysRevB-2006, Baran-PhysRevB-2010}. Phthalocyanines are synthetic macrocyclic compounds, constructed of four 
lobes; each lobe is composed of one pyrrole and one benzene group. Metal-free phthalocyanine (H$_2$Pc) is noted as 
H$_2$C$_{32}$N$_8$H$_{16}$. The flexibility of this molecule provides an opportunity to replace the two central H 
atoms with a metallic atom
($M$Pc) (a metal-oxygen or metal-halogen can also be inserted into the Pc center). This
property has made these molecules very attractive for research, since electronic, optical,
physical and magnetic properties of Pcs can be tuned by changing the metallic center~\cite{Stadler-PhysRevB-2006, 
Baran-PhysRevB-2010, Liao-JChemPhys-2001, Nishi-CS6-2005, Peisert-JPhysChemC-2008}. In
particular, research on transition-metal phthalocyanines has received lots of attention, owing to the significant 
effect of the $d$ orbitals on molecular properties~\cite{Yu-JChemPhys-2012, Petraki-JPhysChemC-2011, 
Petraki-JPhysChemC-2010, Petraki-JPhysChemC-2012, Petraki-JPhysConfSer-2005, Xiao-JPhysCondensMatter-2009, 
Zhang-SurfSci-2005, Peisert-JPhysChemC-2009, Sedona-NatMater-2012, Bartolome-PhysRevB-2010, Lu-JAmChemSoc-1996, 
Gargiani-PhysRevB-2010}. These intrinsically semiconducting molecules, show very high thermal (stable up to 
$500~^{\circ}$C) and chemical stability in addition to their electronic, optical and magnetic properties. These 
characteristics make $M$Pc a promising candidate to be used in organic solar cells~\cite{Riad-ThinSolidFilms-2000, 
delaTorre-ChemCommun-2007, Claessens-ChemRec-2008}, organic light-emitting diodes 
(OLEDs)~\cite{Hung-ApplPhysLett-1999, Gu-ApplPhysLett-1999} and organic field-effect transistors 
(OFETs)\cite{delaTorre-ChemCommun-2007,Yuan-ApplPhysLett-2003}. The function of these devices is influenced by 
another essential factor: interaction and charge transfer at the
interface between $M$Pc and substrate. Numerous studies, applying several experimental and
theoretical methods, have been carried out to investigate the interfacial interaction and charge
transfer, as well as the effect of these interactions on the device functionality. Interaction at
the interface between $M$Pc and substrate determines the layer growth mode, molecular
configuration, charge transfer, magnetic and optical properties of the molecular layer, which
could effectively modify the organic device function. This has encouraged many groups to
investigate the effect of substrate-adsorbate interaction on the properties of the molecular
layer and consequently the potential devices which would have organic components of similar
nature~\cite{Ahmadi-JChemPhys-2012, Yu-JChemPhys-2012, Zhang-SurfSci-2005, Ellis-JApplPhys-2004, 
Naitoh-SurfSci-2001, Takada-JapJApplPhys-2005,Biswas-JChemPhys-2007, Cheng-JPhysChemC-2007, Chizhov-Langmuir-2000, 
Cossaro-JPhysChemB-2004, Evangelista-SurfSci-2003, Fu-PhysRevLett-2007, Gao-PhysRevLett-2007, 
Grzadziel-ThinSolidFilms-2011, Yu-JPhysChemC-2011, Brena-SurfSci-2009}.

Manipulation of the $M$Pc-substrate interaction is possible by either decorating the molecules (adding external 
atoms or molecules to the center or the periphery of 
Pcs)~\cite{Dick-JApplPhys-2005,Evangelista-JPhysChemC-2008,Gerlach-PhysRevB-2005}, by modifying the substrate by 
inserting intermediate layers~\cite{Yu-JPhysChemC-2009,Palmgren-JChemPhys-2008}, or by adding layers on top of 
$M$Pc~\cite{Isvoranu-JPhysCondensMatter-2010, Isvoranu-JChemPhys-2011, Isvoranu-JPhysChemC-2011, 
Isvoranu-JPhysChemC-2011-2, Isvoranu-JChemPhys-2011-2}. For instance, upon adsorption of pyridine on an ordered 
layer of FePc on Au(111),
pyridine molecules coordinate to the iron. This coordination leads to a strong ligand field which modifies the 
magnetic and electronic properties of the iron atom~\cite{Isvoranu-JPhysChemC-2011}. In another study, Gerlach 
\emph{et al.}~\cite{Gerlach-PhysRevB-2005} showed that by fluorination of CuPc adsorbed on Cu(111) and Ag(111), 
the molecular distortion and interfacial interactions are modified. Due to the interaction with the
substrate, CuPc is adsorbed non-planar on these substrates, yet parallel to the sample surface
in both cases. They reported that fluorination resulted in a reduction of the molecule-substrate
attraction and consequently a more ``planar'' molecular configuration, with the molecular layer
further away from the substrate. Bending of $M$Pc molecules on different substrates has been
reported, which generally comes together with an inhomogeneous charge transfer at the
interface between molecule and substrate~\cite{Stadler-PhysRevB-2006, Peisert-JPhysChemC-2008, Yu-JChemPhys-2012, 
Peisert-JPhysChemC-2009, Yamane-PhysRevLett-2010}.

In this article, we have studied the electronic structure of ZnPc on Pt(111) and on two
iodine-induced surface structures, from sub-monolayer to thick films.

\section{Experimental details}

The spectroscopy experiments were done at beamline D1011, at MAX-lab, Swedish national
synchrotron radiation laboratory. D1011 is a bending-magnet beamline which offers photons
in the energy range $40$~eV to $1500$~eV selected by a modified SX-700 plane grating
monochromator~\cite{Nyholm-NIMA-1986}. The experimental system consists of separate analysis and preparation
chambers accessible via a long-travel manipulator.

The photoelectron spectra were measured using a SCIENTA SES200 (upgraded)
electron energy analyzer. The binding energy of all photoelectron spectra is calibrated with
respect to the Fermi level, measured directly on the Pt sample. These spectra are normalized
to the background at the low-binding-energy side of the core-level spectra. The total
experimental resolution for core-level spectra are $180$~meV ($h\nu = 490$~eV, N$1s$), $100$~meV ($h\nu = 382$~eV, 
C$1s$), $20$~meV ($h\nu = 125$~eV, Pt$4f_{7/2}$) and $16$~meV ($h\nu = 110$~eV, I$4d$ and Zn$3d$). The
photoelectron spectra are obtained at normal emission. Numerical curve fitting is done using
Donjiac-\v{S}unji\'{c} line profiles, which includes a Lorentzian broadening ($W_{\text{L}}$) from the finite
core-hole life time, a Gaussian broadening ($W_{\text{G}}$) from limited experimental resolution and
sample inhomogeneities and an asymmetry ($\alpha$) on the high binding energy side due to
excitation of electrons across the Fermi level~\cite{DoniachSunjic-JPhysC-1970}.

For Pt$4f_{7/2}$ best fits were obtained with $W_{\text{L}} = 0.32$~eV, $W_{\text{G}} = (0.30~\text{to}~0.40)$~eV and 
$\alpha = (0.09~\text{to}~0.12)$, while these numbers are $W_{\text{L}} = (0.35~\text{to}~0.45)$~eV, $W_{\text{G}} = 
(0.40~\text{to}~0.55)$~eV and $\alpha = (0.0~\text{to}~0.05)$ for C$1s$ and $W_{\text{L}} = (0.22~\text{to}~0.25)$~eV, 
$W_{\text{G}} = (0.45~\text{to}~0.65)$~eV as well as $\alpha = (0.0~\text{to}~0.05)$ for N$1s$. To fit I$4d$ we use
spin-orbit doublets, with a branching ratio (BR) of $1.2$, a spin-orbit split (SO) of $1.71$~eV, $W_{\text{L}} = 
0.27$~eV, $W_{\text{G}} = (0.28~\text{to}~0.40)$~eV and $\alpha = (0.05~\text{to}~0.12)$.

XAS were collected at the nitrogen K-edge in Total Electron Yield (TEY) mode using
an MCP detector. The photon energies were calibrated using the kinetic-energy difference in
the Pt$4f_{7/2}$ peak measured by first- and second-order light. The absorption spectra were
normalized to the spectrum of a clean sample. XAS were taken at three different angles
between the surface plane and the electric-field vector: $80^{\circ}$, $40^{\circ}$ and $0^{\circ}$.
The preparation chamber is equipped with an ion sputtering gun, low-energy electron
diffraction (LEED) optics, a gas-inlet system and ports for evaporators. The base pressure in
this chamber was lower than $10^{-10}$~mbar. The Pt(111) single crystal was purchased from
\href{http://www.spl.eu/}{Surface Preparation Laboratory, The Netherlands}. Before the measurements, the Pt sample was
prepared by cycles of Ar sputtering, annealing in O$_2$ atmosphere and subsequently flashing at
higher temperature. Ar sputtering was done at $2\times10^{-6}$~mbar, for $20$ minutes; annealing in
oxygen was done at $2\times10^{-6}$~mbar while the sample was heated to $870$~K. After annealing,
the sample was flashed at $1100$~K. A few cleaning cycles were needed to attain a clean and
organized surface. The cleanliness of the sample was confirmed by wide-range PES tracing
any expected impurities and eventually finding none. The clean sample showed a sharp $1\times{}1$
LEED pattern.

Iodine was deposited on the surface from an electrochemical cell. In this cell an AgI
pellet is heated to about $370$~K. An ionic current flows through the cell ($10~\mu$A in this case) and I$_2$
molecules are emitted into the chamber. A $20$-minutes iodine deposition on Pt(111) results in a
$\left(\sqrt{7}\times\sqrt{7}\right)R19.1^{\circ}$ reconstruction, confirmed by LEED and I$4d$ photoelectron spectra. 
Heating this saturated iodine layer, at $430$~K it transforms into a $\left(\sqrt{3}\times\sqrt{3}\right)R30^{\circ}$ 
structure, also confirmed by LEED and I$4d$ core-level spectra.

The ZnPc layers were prepared by sublimation from a quartz crucible with a diameter of
about $5$~mm, after degassing for more than $72$~h. The monolayers of ZnPc were obtained by
deposition on the substrate kept at room temperature. The molecules were purchased from
\href{http://www.sigmaaldrich.com/}{Sigma-Aldrich} ($98$~\% dye content; the remaining $2$~\% consisted mainly of 
water, which was removed by the long outgassing). The thickness of the ZnPc layer was estimated from the
attenuation of the Pt$4f_{7/2}$ signal and the increase of the N$1s$ and C$1s$ signals. At sufficiently
large thickness it is reasonable to apply an exponential attenuation model. The coverage 1~ML
refers to the number of molecules needed to cover the Pt(111) surface, and 6~ML refers to a six times
higher surface coverage, without implying any particular growth mode. We also compare with
other similar systems measured on the same beamline. The method is not exact and the values
we give are therefore approximate.

STM experiments were done in a RHK 3500 UHV STM (in a different system from the
photoemission experiments) using mechanically cut Pt-Ir tips in constant-current mode. The
sample was prepared in a preparation chamber connected to the STM chamber via a gate
valve. This chamber is also equipped with LEED optics, an Ar-ion sputter gun and sample
heating. The sample was mounted on a Mo sample holder. The sample temperature was
measured with chromel-alumel thermocouples, spot-welded on the side of the sample or by a
pyrometer.

\section{Results and discussions}
\subsection{ZnPc on Pt(111)}

In this section the experimental results from ZnPc on Pt(111) are presented. Below,
absorption spectra and core-level photoemission spectra from different steps of the ZnPc
deposition on Pt are shown and the effect of ZnPc adsorption on the substrate, as well as
reactions at the interface and within the molecular layers is discussed. STM images are also
presented from ZnPc layers adsorbed on Pt(111).

\subsubsection{{\rm N K-edge XAS}}

\begin{figure}
\centering
\begin{minipage}[c]{0.493\columnwidth}
\subfloat{\includegraphics[width=\columnwidth]{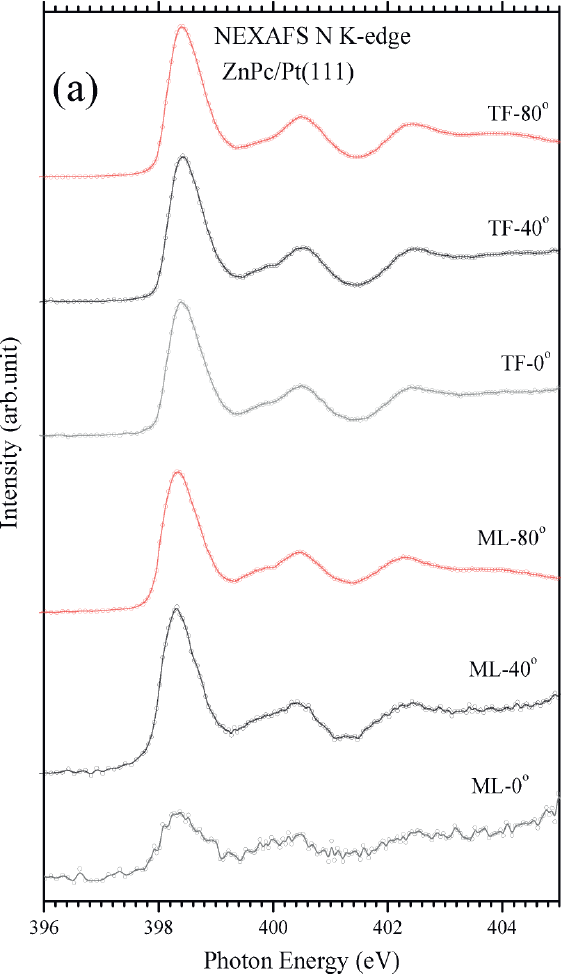}\label{fig:1a}}
\end{minipage}
\hfill
\begin{minipage}[c]{0.493\columnwidth}
\subfloat{\includegraphics[width=\columnwidth]{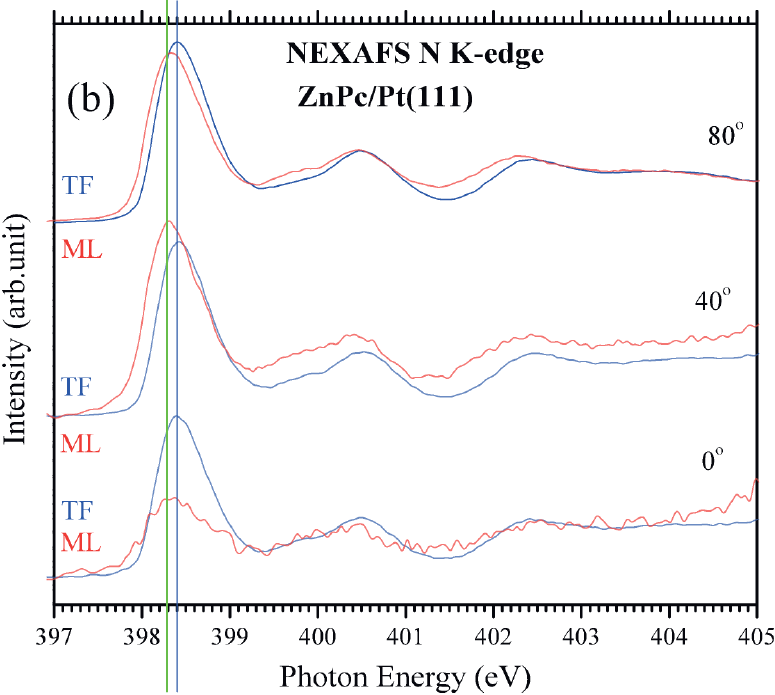}\label{fig:1b}}
\end{minipage}
\caption{(a) N K-edge XAS of a monolayer and a thick film of ZnPc on Pt(111), showing
that ZnPc molecules are almost flat in the ML regime, but tilted at the TF. (b) The magnification of the $\pi^{\ast}$ 
region indicates the participation of nitrogen in the interaction with the Pt substrate at the ML regime.}
\label{fig:1}
\end{figure}

X-ray absorption spectra measured from a monolayer (ML) and a thick film (TF) of ZnPc on
Pt(111) at 3 different angles are presented in Fig.~\ref{fig:1}. A ML refers to the
coverage, in which tightly-packed, flat-lying molecules would cover the entire surface~\cite{Nilson-JPhysChemC-2010}; 
the coverage for a TF here is about 10 MLs. The XAS measurement angles are: $\theta=0^{\circ}$ (normal
incidence), $\theta=40^{\circ}$ and $\theta=80^{\circ}$ (grazing incidence), where $\theta$ is the angle between the 
electric-field vector and the surface plane. In the orbital configuration of phthalocyanines $\pi$ orbitals extend
normal to the molecular plane and $\sigma$ orbitals lie in the molecular plane. In the spectra in Fig.~\ref{fig:1}, 
excitation into the $\pi^{\ast}$ states are seen as peaks between $397$~eV and $403$~eV photon energy,
while excitation into the $\sigma^{\ast}$ states appears at photon energies above 
$404$~eV~\cite{Ahmadi-JChemPhys-2012}. The multiplot demonstrates that molecules are lying almost parallel to Pt(111) 
in the monolayer, while they are slightly tilted in the thick film.

A closer look at the N K-edge spectra in Fig.~\ref{fig:1b} discloses a shoulder-like peak on
the lower photon energy side of the first resonance associated to excitation into the LUMO in
the ML spectra [marked in Fig.~\ref{fig:1b}], especially visible in the spectra measured at $\theta=40^{\circ}$
and $\theta=80^{\circ}$. Another noticeable change is the variation in the relative intensity of the
peaks in each spectrum going from the ML to the TF. A similar behavior was observed for FePc adsorbed on Ag(111) but 
interestingly not for FePc on Au(100)~\cite{Petraki-JPhysChemC-2012}. A splitting of the first resonance is caused by a 
hybridization of nitrogen orbitals and metal-related states.

\subsubsection{{\rm Pt$4f_{7/2}$ photoemission}}

\begin{figure}
\centering
\includegraphics[width=0.6\columnwidth]{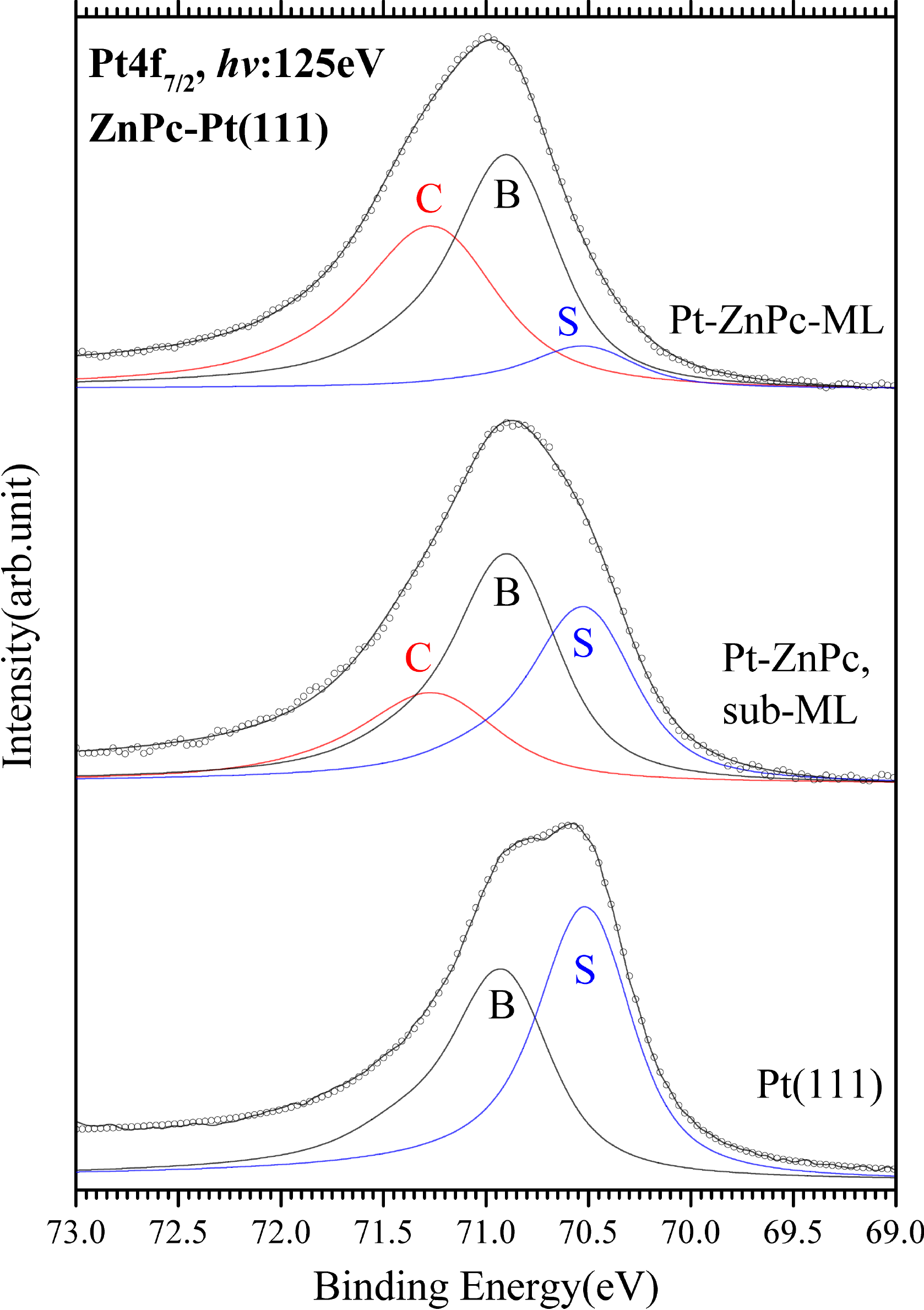}
\caption{Pt$4f_{7/2}$ spectra recorded from Pt(111)-ZnPc after different preparations.}
\label{fig:2}
\end{figure}

In Fig.~\ref{fig:2}, Pt$4f_{7/2}$ photoemission spectra from different sample preparations are presented.
The lowermost spectrum is measured from clean Pt(111) and comprises two peaks: the bulk
peak at $70.91$~eV and the surface peak located at a $0.41$~eV lower BE, in good agreement with
previous studies~\cite{Puglia-SurfSci-1995, Janin-SurfSci-2001, Zhu-SurfSci-2003}. Upon adsorption of ZnPc the 
surface peak loses intensity drastically and
is further reduced at higher coverage, unlike the ZnPc/Au(111) system, where even at a high
coverage, the surface peak keeps its intensity~\cite{Ahmadi-JChemPhys-2012}. Another difference compared to Au(111) is
that after adsorption of ZnPc on Pt(111), a new peak (C) appears at higher BE. The
appearance of the new chemically shifted peak at $E_{\text{C}}$ following adsorption of organic
molecules has been reported previously; e.g. for propene and ethylidine: $\Delta{}E_{\text{C}} = E_{\text{C}} - 
E_{\text{B}} = 0.26$~eV, 2-butenal: $\Delta{}E_{\text{C}} = 0.27$~eV and CCH$_3$: $\Delta{}E_{\text{C}} = 0.36$~eV are 
observed~\cite{Janin-SurfSci-2001, Bjoerneholm-SurfSci-1994}. Adsorption of other molecules such as 
CO~\cite{Bjoerneholm-SurfSci-1994} and O$_2$~\cite{Puglia-SurfSci-1995} also resulted in adsorption-induced chemical 
shifts up to $1$~eV, depending on the coverage. Our observation indicates that ZnPc is chemisorbed on Pt(111). The total 
integrated area under S and C remains roughly at a value which is
comparable with the area under the surface peak of the clean surface. Even at the highest
coverage, S has $9$~\% relative intensity, meaning that some surface atoms are untouched by the
adsorbate layer. The initial surface peak represents one monolayer of Pt, contributing to $52$~\%
of the total Pt$4f_{7/2}$ intensity. Thus, $9$~\% represents $0.17$ ML of surface atoms.
Kr\"oger \emph{et al.} showed that the $D4h$ symmetry of the molecule is responsible for the vanishing intrinsic 
electrostatic moment~\cite{Kroeger-NewJPhys-2010}, giving weak intermolecular forces where the
favorable adsorption positions are determined mainly by the interaction of molecules with the
substrate~\cite{Kroeger-NewJPhys-2010, Cheng-JPhysChemC-2007}. However, in our case the chemisorption bond will lead to 
local geometric and/or electronic ``deformation'' of both substrate and molecule, which as previously shown will create 
a substrate-mediated intermolecular repulsion~\cite{Kroeger-NewJPhys-2010}.

\begin{table}
\caption{\label{tab:I} Pt$4f$ curve-fitting results.}
\begin{ruledtabular}
\begin{tabular}{cccccc}
 & $\Delta{}E_\text{S}$/eV & $I_\text{S}$/\% & BE$_\text{B}$/eV & $\Delta{}E_\text{C}$/eV & 
$I_\text{C}$/\% \tabularnewline
\noalign{\smallskip}
\hline
\noalign{\smallskip}
Pt(111) & $0.41$ & $52$ & $70.91$ & -- & -- \tabularnewline
Pt-ZnPc-subML & $0.37$ & $34$ & $70.91$ & $0.36$ & $22$ \tabularnewline
Pt-ZnPc-ML & $0.37$ & $9$ & $70.91$ & $0.36$ & $43$ \tabularnewline
\end{tabular}
\end{ruledtabular}
\end{table}

\subsubsection{{\rm STM}}

\begin{figure}
\centering
\subfloat{\label{fig:3a}}
\subfloat{\label{fig:3b}}
\subfloat{\label{fig:3c}}
\subfloat{\label{fig:3d}}
\includegraphics[width=\columnwidth]{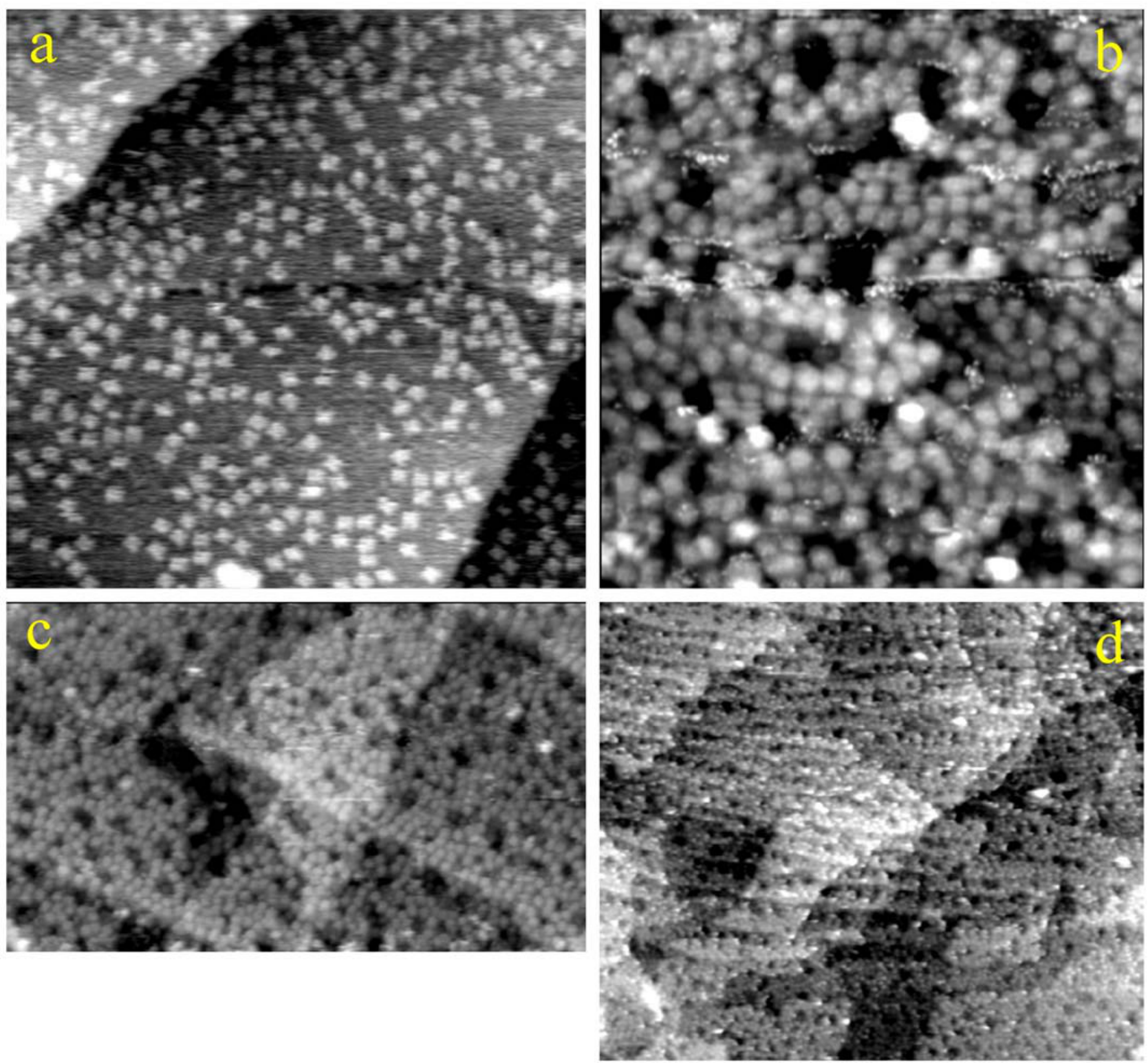}
\caption{Room-temperature STM images of ZnPc on Pt(111). (a) Sub-ML of ZnPc/Pt(111); $360\times{}360$~\AA{}$^{2}$, 
$60$~pA, $125$~mV, and multilayer of ZnPc/Pt(111): (b) $330\times{}350$~\AA{}$^{2}$, $68$~pA, $9$~mV, (c) 
$900\times{}540$~\AA{}$^{2}$, $68$~pA, $125$~mV (d) $1800\times{}1600$~\AA{}$^2$, $0.8$~nA, $53$~mV.}
\label{fig:3}
\end{figure}

In Fig.~\ref{fig:3} we present STM images from ZnPc-Pt(111) at sub-monolayer coverage and
from a thin film. Figure~\ref{fig:3a} illustrates that ZnPc molecules are lying flat on the surface and
are not tightly packed; instead they are sitting far from each other in a scattered manner. A
significant observation is that the molecular arrangements are different; in some areas, a few
molecules are aligned in one orientation but the arrangement is very local and in connection
with the specific region of the substrate. This indicates the determining role of the substrate-molecule interaction and 
confirms the relatively weaker molecule-molecule interaction mentioned 
above~\cite{Cheng-JPhysChemC-2007,Cheng-JPhysChemC-2007-2,Zhang-JPhysChemC-2011}. Figures~\ref{fig:3b} to~\ref{fig:3d} 
show images from a multilayer ZnPc film. Clearly,
single molecules are resolved in Fig.~\ref{fig:3b}. ZnPc appears to lie with the molecular plane
parallel to the surface, and with a preferred side-to-side alignment to neighboring molecules.
However, this ordering does not extend over more than a few molecules. Although the
molecular packing is denser, a complete layer is not formed. Instead nanometer-sized holes
appear across the surface. Figure~\ref{fig:3c} is a $900\times{}540$~\AA{}$^{2}$ area, which shows that actually some
repeated order exists in a larger view over the surface. Figure~\ref{fig:3d} demonstrates an even larger
area over the same region ($1800\times{}1600$~\AA{}$^{2}$), confirming the repetitive pattern. The relative
area of the holes is on the order of $10$~\% to $20$~\%, in good agreement with the Pt$4f$ results,
pointing to uncovered Pt in the bottom of the holes.

A similar porous molecular arrangement was observed for FePc on Au(111) at a so-called higher sub-monolayer coverage, 
particularly at the elbows of the herringbone structure~\cite{Cheng-JPhysChemC-2007-2}. Cheng \emph{et al.} explained 
the origin of this ring-like structure to be the directional attraction between 
molecules~\cite{Cheng-JPhysChemC-2007-2}, driven by fitting the benzene group of one molecule to the hollow site of its 
neighbor. In this position hydrogen atoms of benzene are placed close to nitrogen atoms of the neighboring molecule. 
Since the nitrogen atom has an unshared pair of electrons and
the hydrogen atom possesses a net positive charge, this position is favorable and motivates this ring-like molecular 
arrangement~\cite{Cheng-JPhysChemC-2007-2}.

\subsubsection{{\rm Molecular core-level spectroscopy}}

\begin{figure*}
\centering
\begin{minipage}[c]{0.28\textwidth}
\subfloat{\includegraphics[width=\textwidth]{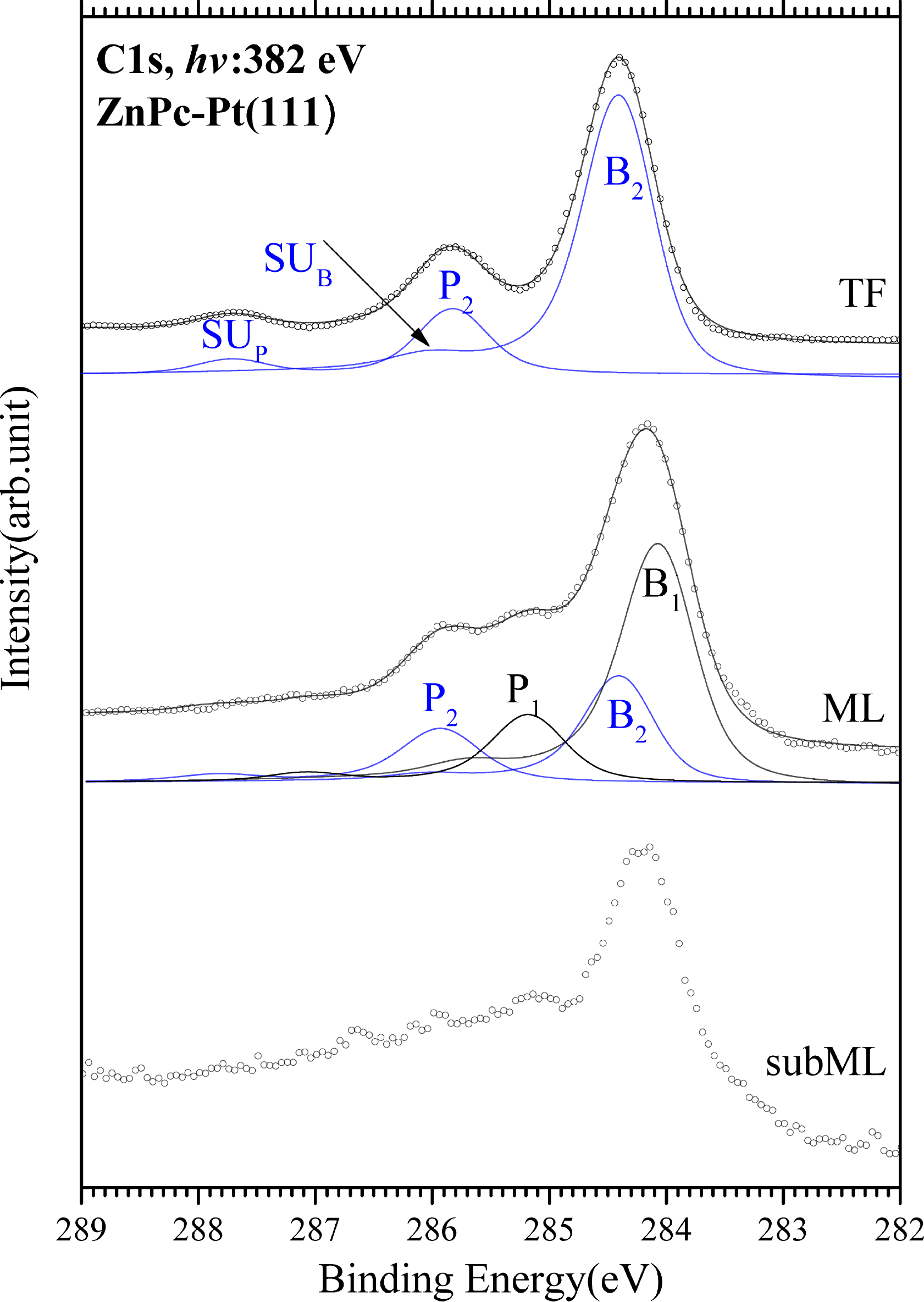}\label{fig:4a}}
\end{minipage}
\hspace*{2mm}
\begin{minipage}[c]{0.28\textwidth}
\subfloat{\includegraphics[width=\textwidth]{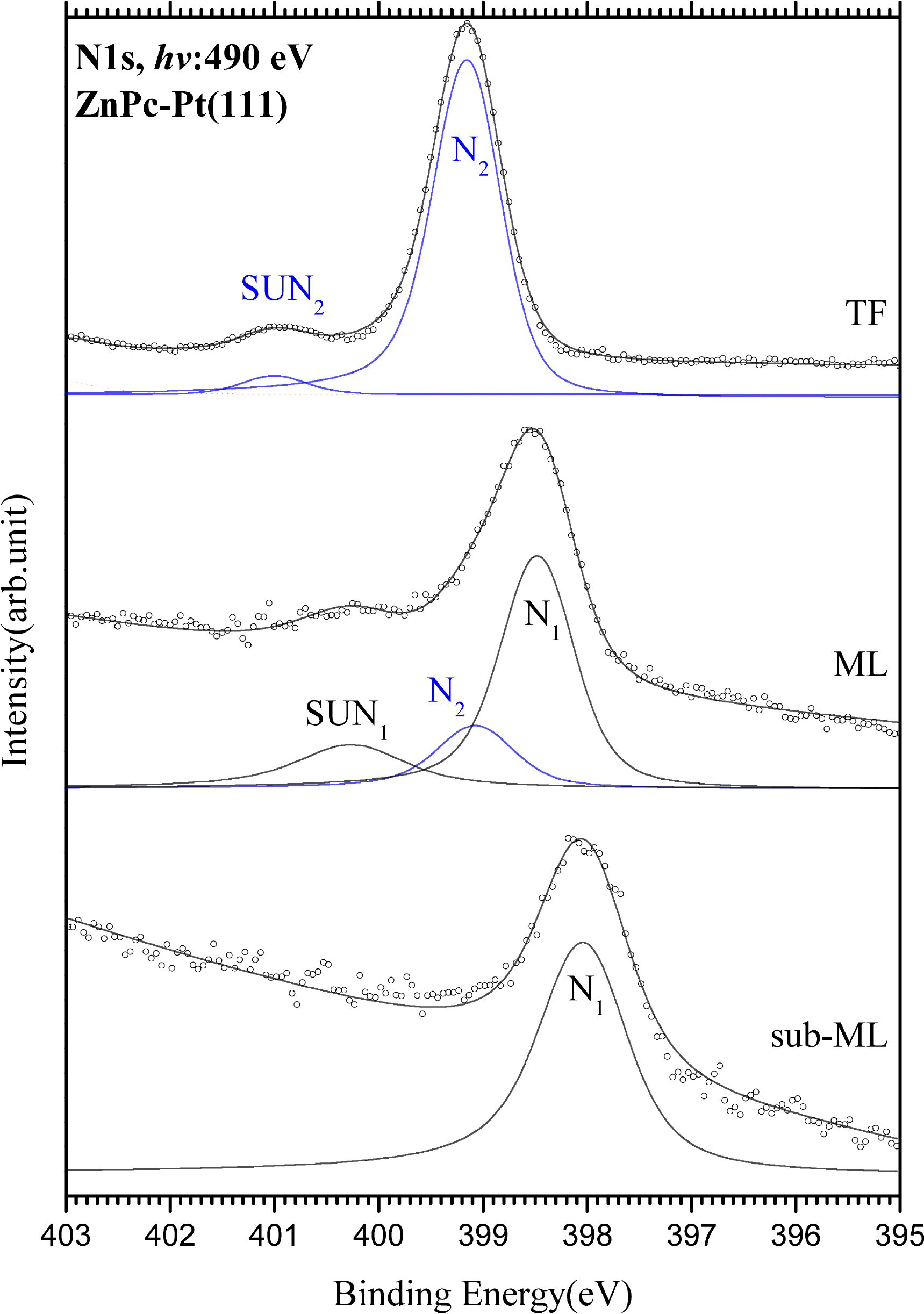}\label{fig:4b}}
\end{minipage}
\hspace*{2mm}
\begin{minipage}[c]{0.275\textwidth}
\subfloat{\includegraphics[width=\textwidth]{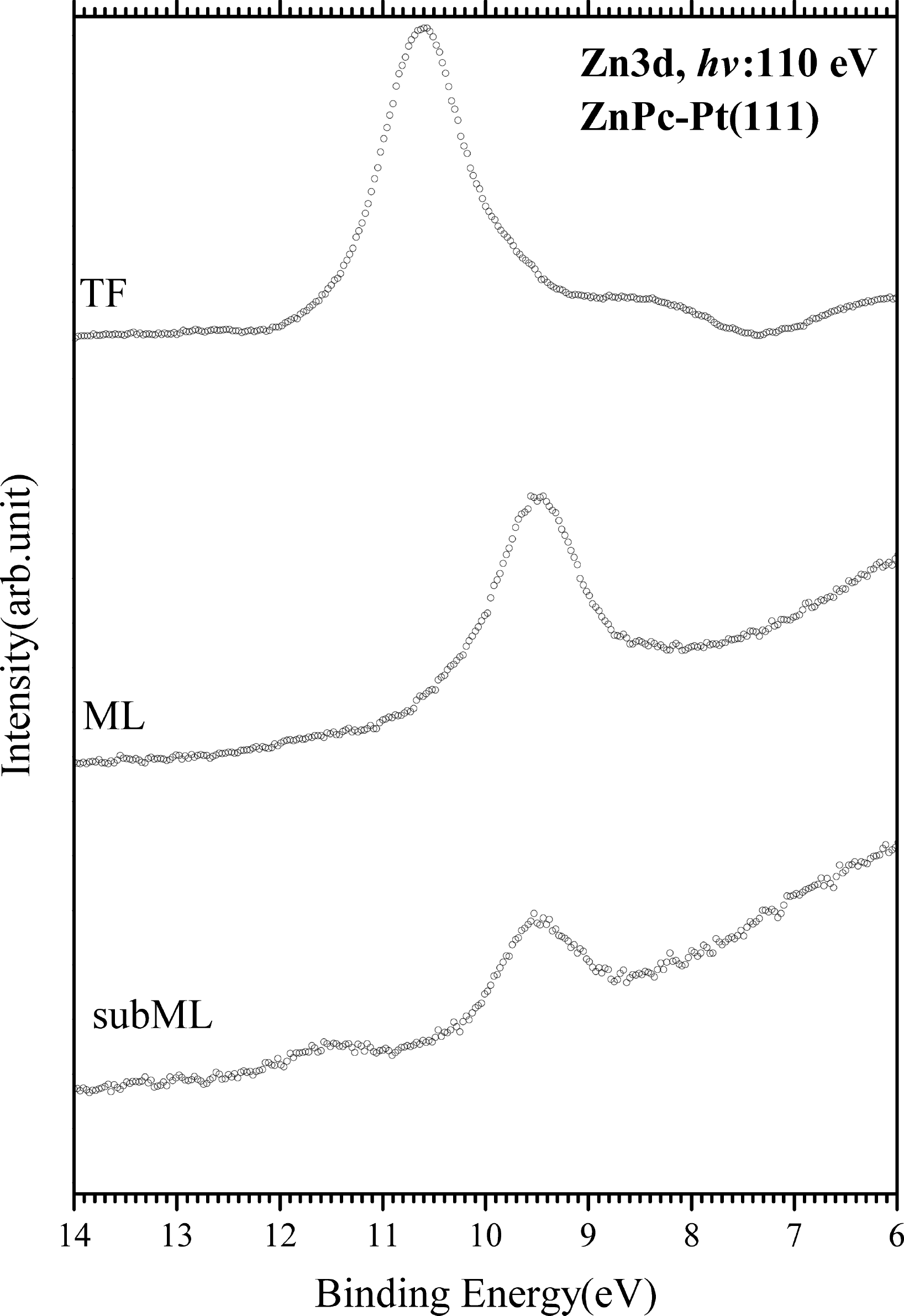}\label{fig:4c}}
\end{minipage}
\caption{(a) C$1s$, (b) N$1s$ and (c) Zn$3d$ photoemission spectra of ZnPc on Pt(111).}
\label{fig:4}
\end{figure*}

As mentioned in previous sub-sections, the ZnPc-Pt interaction leads to a reduction
of the surface-shifted photoemission peak intensity and the appearance of an extra Pt$4f_{7/2}$ peak
at higher binding energy. The binding energy of this chemically shifted peak (C) indicates a
charge transfer from Pt to ZnPc. Here, we use the molecular-core-level spectra, measured on a
ML and a TF of ZnPc and compare with the results of ZnPc on Au(111) to investigate the
effect of the substrate on the molecular core levels in order to reveal the characteristics of the
interaction at the interface.

C$1s$, N$1s$ and Zn$3d$ photoemission spectra are shown in Fig.~\ref{fig:4} as a function of
molecular layer thickness. In Fig.~\ref{fig:4a}, C$1s$ spectra are presented. The thick-film spectrum
shows a molecular-like C$1s$ spectrum consisting of three peaks: C$_\text{B}$ at $284.40$~eV (benzene
carbon), the second peak contains C$_{\text{P}}$ (pyrrole carbon) and the shake-up of the benzene peak
(SU$_\text{B}$); C$_\text{P}$ is centered at $285.80$~eV, and SU$_\text{B}$ and SU$_\text{P}$ (shake-up structure of 
the pyrrole carbon) at $286.03$~eV and $287.72$~eV, respectively. These peaks are located at the same BE as
for the thick film deposited on Au(111)~\cite{Ahmadi-JChemPhys-2012}.

It is clear that C$1s$ spectra are not molecular-like at low coverage; ZnPc is disturbed by
the interaction with Pt. This is not the case when adsorbed on Au 
surfaces~\cite{Ahmadi-JChemPhys-2012,Peisert-JPhysChemC-2008,Petraki-JPhysChemC-2010,Peisert-JPhysChemC-2009,
Cossaro-JPhysChemB-2004,Shariati-JPhysChemC-2013,Evangelista-JChemPhys-2009, Peisert-JApplPhys-2002, Pi-JApplPhys-2009},
which are not as reactive as Pt. Yet, on Ag, Cu and TiO$_2$ surfaces, C$1s$ spectra are also modified at sub-monolayer 
thicknesses~\cite{Yu-JChemPhys-2012,Yamane-PhysRevLett-2010,Ruocco-JPhysChemC-2008}. Our curve fitting shows that at ML 
C$1s$ needs
components representing carbon at the interface and carbon in the second layer. There are two
C$_\text{B}$ (benzene carbon) peaks and two C$_\text{P}$ (pyrrole carbon) peaks: interface benzene (B$_1$) at
$284.08$~eV and second-layer benzene (B$_2$), at $284.41$~eV. Interface pyrrole (P$_1$) is at $285.17$~eV
and second layer pyrrole (P$_2$) at $285.92$~eV. Thus, C$_\text{P}$ shifts more than C$_\text{B}$, i.e. pyrrole carbon
are more affected by the interfacial interaction than benzene carbon. This site-specific
coverage-dependent shifts have been observed before for MgPc on polycrystalline gold,
where by increasing the thickness from $0.5$~nm to $9$~nm, C$_\text{B}$ shifted $0.4$~eV to higher BE, while
C$_\text{P}$ shifted $0.6$~eV in the same direction~\cite{Peisert-JPhysChemC-2008}. Also in ZnPc/TiO$_2$, C$_\text{P}$ 
shifted $1.5$~eV to higher BE in a thick film with respect to a ML, but C$_\text{B}$ shifted only 
$0.8$~eV~\cite{Yu-JChemPhys-2012}. The C$_\text{B}$ peak of the ZnPc ML is centered at $284.08$~eV which is at $0.47$~eV 
higher BE than for a ZnPc ML adsorbed on
Au(111) (this is $0.18$~eV for C$_\text{P}$)~\cite{Ahmadi-JChemPhys-2012}. This indicates a lower electron density 
around the carbon
atoms at the interface between ZnPc and Pt(111) than Au(111). There is also a shift to higher
BE going from a ML to a TF ($0.32$~eV for C$_\text{B}$ and $0.75$~eV for C$_\text{P}$). Generally, this shift is
explained as weaker core-hole screening through the substrate at thick films with respect to
low-coverage films. This shift also depends on the charge transfer at the interface, i.e. in
systems where molecules receive electrons from the substrate, by increasing the coverage the
molecules are further from the substrate, the core-level binding energies increase in the thick
film~\cite{Kroeger-NewJPhys-2010}. Telling these two sources of a BE shift apart is not straightforward and requires 
more data analysis; this will be discussed in more detail later in this section. Another noticeable
observation is that the binding energies of B$_2$ and P$_2$ are the same as C$_\text{B}$ and C$_\text{P}$ at the thick
film, indicating that the molecules are practically decoupled from the substrate already at the
second layer (at least the carbon rings). Previously it was claimed that CuPc on Ag(111) was decoupled from the 
substrate from the third layer (measured at $80$~K)~\cite{Kroeger-NewJPhys-2010}.

N$1s$ spectra of ZnPc on Pt(111) are shown in Fig.~\ref{fig:4b}. The thick-film spectrum is a
single peak located at $399.16$~eV which represents two different groups of nitrogen atoms in
the Pc molecule; the separation of these two peaks is too small to be resolved in the 
spectrum~\cite{Aalund-JChemPhys-2006}. The shake-up structure is located at $1.82$~eV higher BE, the same as observed 
for ZnPc/TiO$_2$~\cite{Yu-JChemPhys-2012} and ZnPc/Au(111)~\cite{Ahmadi-JChemPhys-2012}. The first two spectra at the 
bottom are from a sub-ML and ML film and their main peaks are broader than in the thick-film spectrum. The curve-fitting 
results show that in the ML region, two peaks (with the same line profile as in the TF spectrum) are present: N$_1$ at 
lower BE and N$_2$ at higher BE. Since it seems that in the TF, only
the higher BE peak is present, we ascribe N$_1$ to the nitrogen atoms directly in contact with the
Pt surface and N$_2$ to the ones in the second layer. This is in agreement with the suggested
charge-transfer (CT) direction at the interface of ZnPc and Pt(111).

The N$1s$ binding energy for a ML is $0.60$~eV higher than for a ML of ZnPc on Au.
This shift is larger than the corresponding C$_\text{B}$ and C$_\text{P}$ shifts. Another observation is that N$1s$
shifts $0.65$~eV to higher binding energy by increasing the coverage from ML to TF, which is
larger than observed for C$_\text{B}$ ($0.33$~eV) but close to the value for C$_\text{P}$ ($0.75$~eV). Based on the
explanations given earlier for C$1s$, either nitrogen atoms suffer more from poor core-hole
screening in the thick film or the charge transfers at the interface between Pt-C$_\text{B}$ and Pt-N are
different in the monolayer region.

In order to get a clearer image of the charge transfer between ZnPc and Pt and how
the different atoms are affected by this interaction, we present and discuss core-level spectra
from the central metal atom. Zn$3d$ spectra are shown in Fig.~\ref{fig:4c}. The Zn$3d$ peak shows up
at $9.46$~eV at sub-monolayer coverage and $9.50$~eV for a monolayer, which is at lower BE than ZnPc on Au(111) 
($9.82$~eV)~\cite{Ahmadi-JChemPhys-2012}. As a result of further ZnPc adsorption, a component at higher BE of the first 
peak appears in the Zn$3d$ spectrum. At first, by increasing the coverage to 1 ML, the low-BE peak grows but at even 
higher coverage, the high-BE component becomes dominant and in the molecular thick film only the one at higher BE 
exists. Thus, we
conclude that the low-BE component stems from molecules at the interface, while the high-BE component has contributions 
from molecules without direct contact with the substrate. The binding-energy shift between these two peaks is $0.6$~eV 
for all preparations in which they coexist. In the thick film, the interfacial component is absent, since there is no 
or only a very weak contribution from the interface to the photoemission signal of the thick film. A lower
binding energy of the interfacial component compared to the molecular-like one is due to the
availability of more electrons at the interface. The energy difference between the low-BE
peak of ML and the TF peak ($1.1$~eV) is larger than the energy shift observed for N ($0.65$~eV),
C$_\text{P}$ ($0.75$~eV) and C$_\text{B}$ ($0.32$~eV). This means that the interfacial interaction has the greatest
effect on the Zn. The TF peak experiences a shift to a higher BE compared to the high-BE
peak of the ML region, implying that Zn, unlike C and N is not decoupled from the substrate
at this coverage. Furthermore, the comparison of core-level binding energies on Pt(111) and
Au(111) reveals that C$1s$ and N$1s$ have higher binding energies on Pt while for the Zn$3d$
spectra, the binding energy for a ML on Pt is lower than on Au(111). In other words, the
inhomogeneous charge distribution is more pronounced when ZnPc is adsorbed on Pt.
The difference in coverage-dependent shifts for different components can be
comprehended by several parameters affecting the energy shifts. It can be due to either local
or global effects. Since a global effect is expected to influence the shifts for different
components almost equally, a more local initial or/and final state effect should be responsible
for the diverse BE shifts in this system. One possible explanation is different screening for
different atoms. Peisert \emph{et al.} showed that for ZnPc (and MgPc) adsorbed on Au(100),
polarization screening is not enough and contributions from charge-transfer screening have to be 
considered~\cite{Peisert-JPhysChemC-2009,Kolacyak-ApplPhysA-2009}. Charge-transfer screening is more local compared to 
polarization screening. An inhomogeneous charge transfer for $M$Pc molecules has been reported
before~\cite{Peisert-JPhysChemC-2008,Yu-JChemPhys-2012,Peisert-JPhysChemC-2009,Yamane-PhysRevLett-2010}. Peisert 
\emph{et al.} reported different shifts between different atoms when MgPc was
adsorbed on Au(111): $-0.6$~eV for Mg as well as C$_\text{P}$ and $-0.4$~eV for C$_\text{B}$ and N. They suggested
that this inhomogeneity is a result of different molecule-substrate distances and/or a site-dependent wave function 
overlap between the metal and the organic molecule~\cite{Peisert-JPhysChemC-2008}. Both these effects influence the 
charge transfer time scales.

A XSW (X-ray standing wave) and ARPES (angle-resolved photoemission spectroscopy) study of ZnPc on Cu(111) 
demonstrated site-specific molecule-substrate interactions~\cite{Yamane-PhysRevLett-2010} and showed that ZnPc is 
distorted on Cu(111), i.e., the Zn atom is pulled down toward the substrate while the organic rings are located farther 
from the substrate. They observed different thickness-dependent shifts for different atoms due to a change of the
molecular geometry and the atom-substrate distances. Interestingly, by fluorination of ZnPc,
they observed an increase in the molecular-layer distance to the substrate followed by a
decrease in height variations for different atoms of ZnPc (a decrease of the protrusion of Zn)
and consequently smaller energy shifts~\cite{Yamane-PhysRevLett-2010}. Thus, based on literature and our own 
observations, it is very probable that ZnPc is also distorted on Pt. A bending of ZnPc, in a way that the Zn
atom is pulled down closer to the Pt surface and the organic rings are bent upwards, together
with the fact that Pt donates electrons to ZnPc, would explain the difference in the coverage-
dependent shifts as well as the lower BE of Zn$3d$ and the higher BE of C$1s$ and N$1s$ on this
surface compared to Au(111). The Zn and C$_\text{P}$ atoms experience a different charge transfer
than the C$_\text{B}$ and N atoms. The details of the different thickness-dependent BE shifts (from our
results and references) are given in Table~\ref{tab:II}.

\begin{table}
\caption{\label{tab:II} Coverage-dependent core-level shifts between a ML and a TF of $M$Pc adsorbed on different
substrates. The numerical values are given in units of electronvolts.}
\begin{ruledtabular}
\begin{tabular}{ccccc}
 & C$_\text{B}$ & C$_\text{P}$ & N & $M$ \tabularnewline
\noalign{\smallskip}
\hline
\noalign{\smallskip}
ZnPc/TiO$_2$~\cite{Yu-JChemPhys-2012} & $0.80$ & $1.5$ & $1.7$ & -- \tabularnewline
FePc/TiO$_2$~\cite{Palmgren-JChemPhys-2008} & $1.2$ & $1.1$ & $1.3$ & -- \tabularnewline
ZnPc/Au~\cite{Ahmadi-JChemPhys-2012} & $0.97$ & $0.99$ & $0.7$ & $1.0$ \tabularnewline
FePc/Au~\cite{Ahmadi-JChemPhys-2012} & $0.40$ & $0.70$ & $0.20$ & -- \tabularnewline
ZnPc/Pt & $0.32$ & $0.75$ & $0.65$ & $1.1$ \tabularnewline
MgPc/Au~\cite{Peisert-JPhysChemC-2008} & $0.40$ & $0.60$ & $0.40$ & $0.60$ \tabularnewline
ZnPc/PtI-$\left(\sqrt{3}\times\sqrt{3}\right)R30^{\circ}$ & $0.12$ & $0.12$ & $0.08$ & $0.10$ \tabularnewline
ZnPc/PtI-$\left(\sqrt{7}\times\sqrt{7}\right)R19.1^{\circ}$ & $0.51$ & $0.48$ & $0.41$ & $0.44$ \tabularnewline
\end{tabular}
\end{ruledtabular}
\end{table}

Our results do not provide sufficient evidence that a bending of the molecular plane is
the only reason for the site-dependent charge distribution in the molecules. Hence, we also
discuss the other possibility, the effect of different wave-function overlap for the different
atoms of the molecule with the substrate. Our Pt$4f_{7/2}$ and Zn$3d$ photoemission results confirm
Zn-Pt interaction at the interface. Moreover, XAS displayed that nitrogen atoms are involved
in the interaction between the molecules and the substrate. This is in agreement with the
results from the FePc/Ag(111) system, where Fe L-edge and N K-edge XAS confirmed that both Fe and N atoms are involved 
in the interfacial interaction~\cite{Petraki-JPhysChemC-2012}. Interestingly, the splitting of
the first resonance and the change of the line shape were not observed for FePc/Au(100),
indicating that N is less involved in the bond to Au than to Ag~\cite{Petraki-JPhysChemC-2012}. Other studies are in 
line with this; for FePc on Au(111)~\cite{Ahmadi-JChemPhys-2012,Isvoranu-JChemPhys-2011} and on 
Au(110)~\cite{Betti-JPhysChemC-2010,Gargiani-PhysRevB-2010}, the metal $d$-states on the central atom are the
main molecular contributor to the molecule-surface bond.

Altogether, our core-level-spectroscopy results suggest a non-planar molecular
configuration for ZnPc on Pt(111), induced by an attractive Zn-Pt interaction. It is shown that
ZnPc molecules do not form a complete monolayer on a Pt surface. Instead, upon increasing
the coverage (up to 1 ML), a surface-mediated intermolecular repulsion forces ZnPc into
multilayers. The molecules are lying almost parallel to the substrate at lower coverage, while
they prefer a tilted position for a higher coverage, due to the decrease of the molecule-substrate interactions. Our 
photoemission results illustrate that ZnPc is practically decoupled from Pt, starting from the second layer.

\subsection{ZnPc on Pt-I}

Adsorption of iodine on Pt(111) leads to two different reconstructions: 
$\left(\sqrt{3}\times\sqrt{3}\right)R30^{\circ}$ at 1/3 ML and $\left(\sqrt{7}\times\sqrt{7}\right)R19.1^{\circ}$ at 
3/7 ML iodine coverage~\cite{DiCenzo-PhysRevB-1984}. In the case of $\left(\sqrt{3}\times\sqrt{3}\right)$, iodine 
occupies three-fold hollow sites~\cite{Tkatchenko-SurfSci-2005}. For the $\left(\sqrt{7}\times\sqrt{7}\right)$ 
reconstruction, both fcc and hcp hollow sites together with the top sites are occupied~\cite{Jo-SurfSci-1992}.

\subsubsection{{\rm Pt$4f_{7/2}$}}

\begin{figure}
\centering
\begin{minipage}[c]{0.493\columnwidth}
\subfloat{\includegraphics[width=\columnwidth]{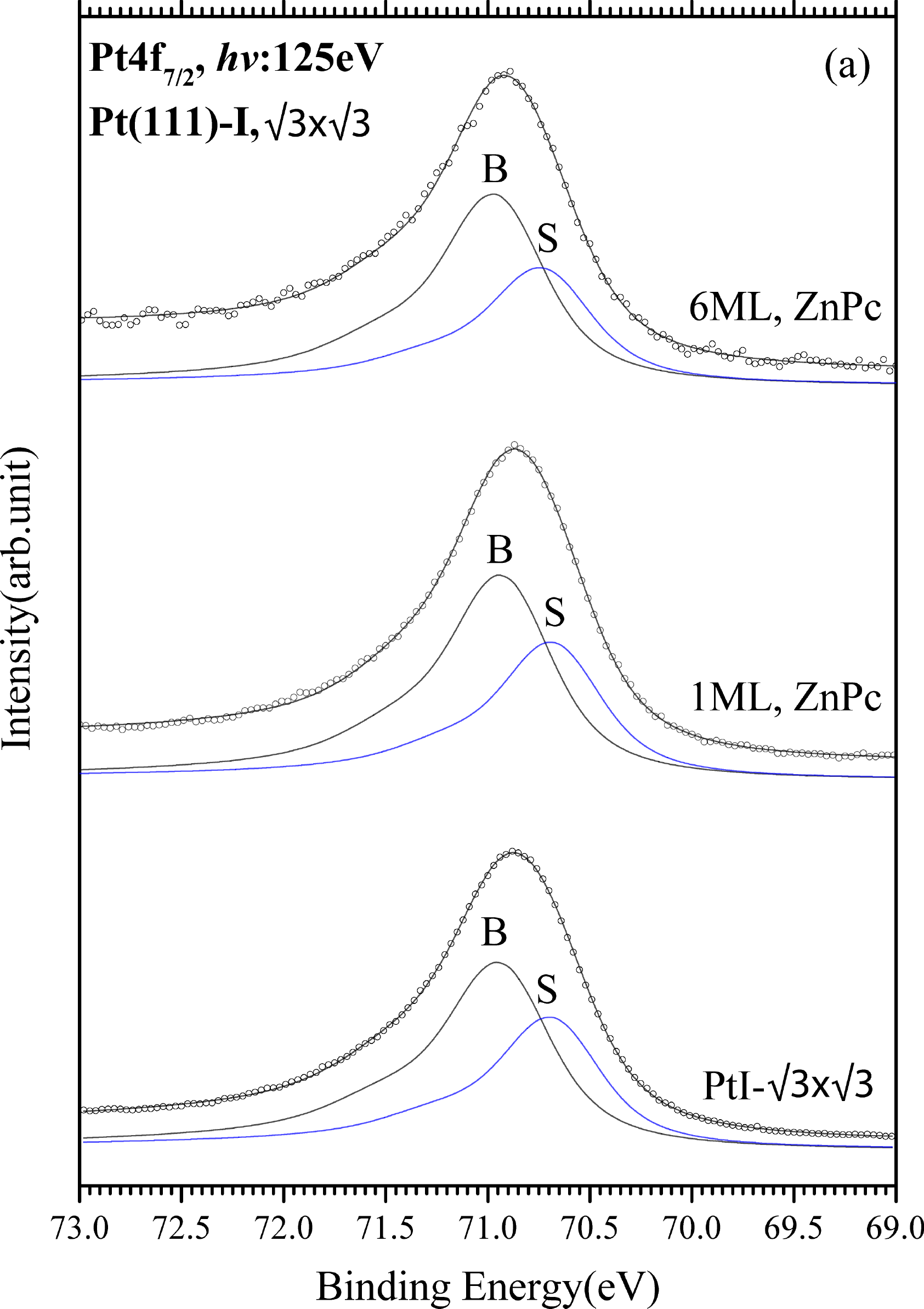}\label{fig:5a}}
\end{minipage}
\hfill
\begin{minipage}[c]{0.493\columnwidth}
\subfloat{\includegraphics[width=\columnwidth]{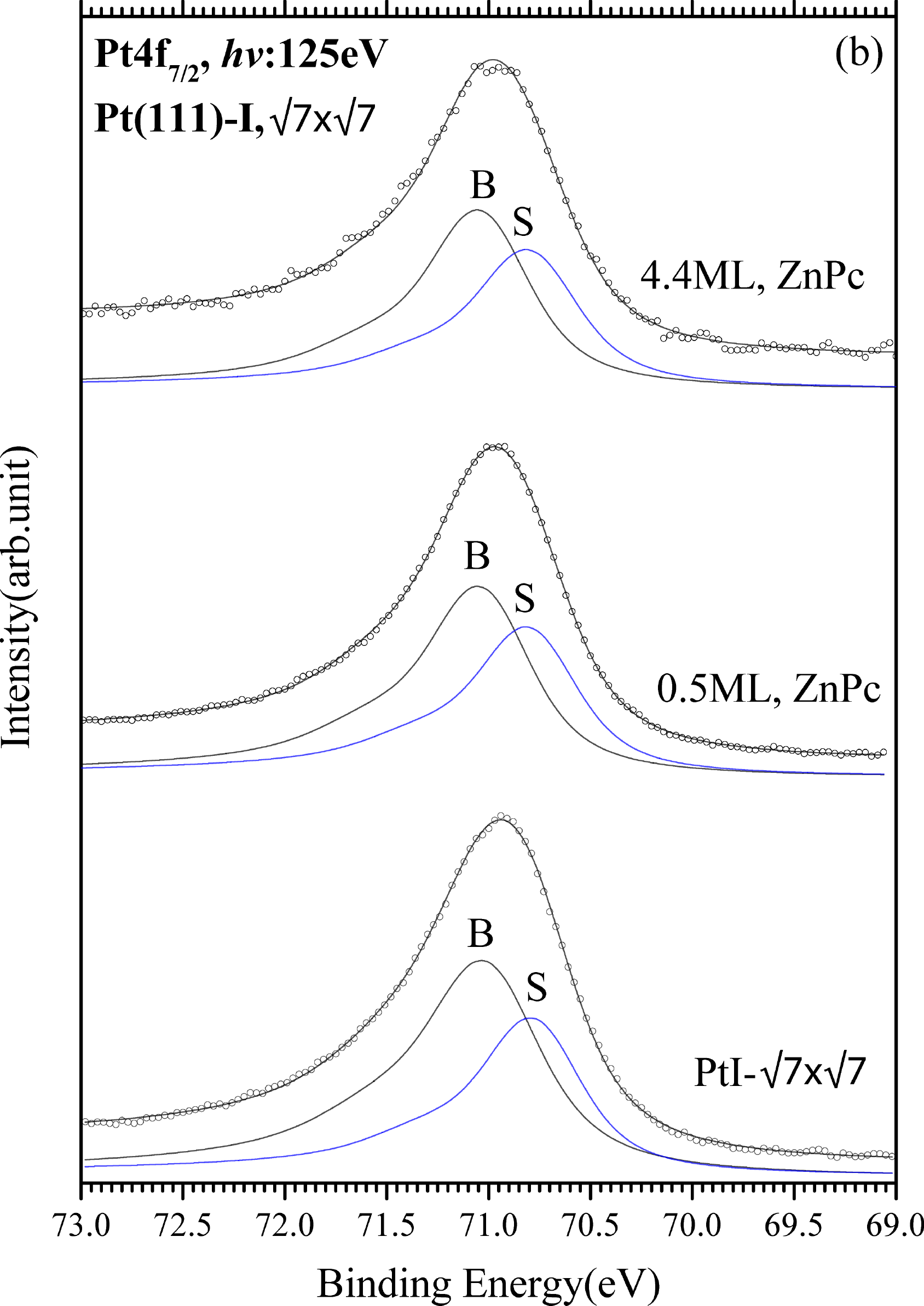}\label{fig:5b}}
\end{minipage}
\caption{Pt$4f_{7/2}$ spectra of ZnPc on (a) PtI-$\left(\sqrt{3}\times\sqrt{3}\right)R30^{\circ}$ and (b) 
PtI-$\left(\sqrt{7}\times\sqrt{7}\right)R19.1^{\circ}$.}
\label{fig:5}
\end{figure}

Pt$4f_{7/2}$ photoemission spectra from the Pt(111)-I surfaces together with their numerical fits are
shown in Fig.~\ref{fig:5}. These spectra were measured before and after ZnPc adsorption, as
indicated in the figure. The bottom spectrum in Fig.~\ref{fig:5a}, from $\left(\sqrt{3}\times\sqrt{3}\right)$ before 
ZnPc deposition, consists of two peaks: the bulk peak at $70.92$~eV and the surface-induced peak at
$70.67$~eV. The surface shift is $0.25$~eV. The relative surface intensity is $42$~\%. For the 
$\left(\sqrt{7}\times\sqrt{7}\right)$ surface the spectrum is practically the same despite the different surface order 
and different adsorption geometries; the surface shift is $0.24$~eV and the relative intensity is $42$~\%.
The top spectra in each panel were recorded after adsorption of ZnPc. There are no observable
changes in either of the spectra. The surface shifts and the relative surface intensity remain
unaltered. This shows that the ZnPc does not affect Pt when an iodine layer is present.

\subsubsection{{\rm XAS}}

\begin{figure}
\centering
\begin{minipage}[c]{0.493\columnwidth}
\subfloat{\includegraphics[width=\columnwidth]{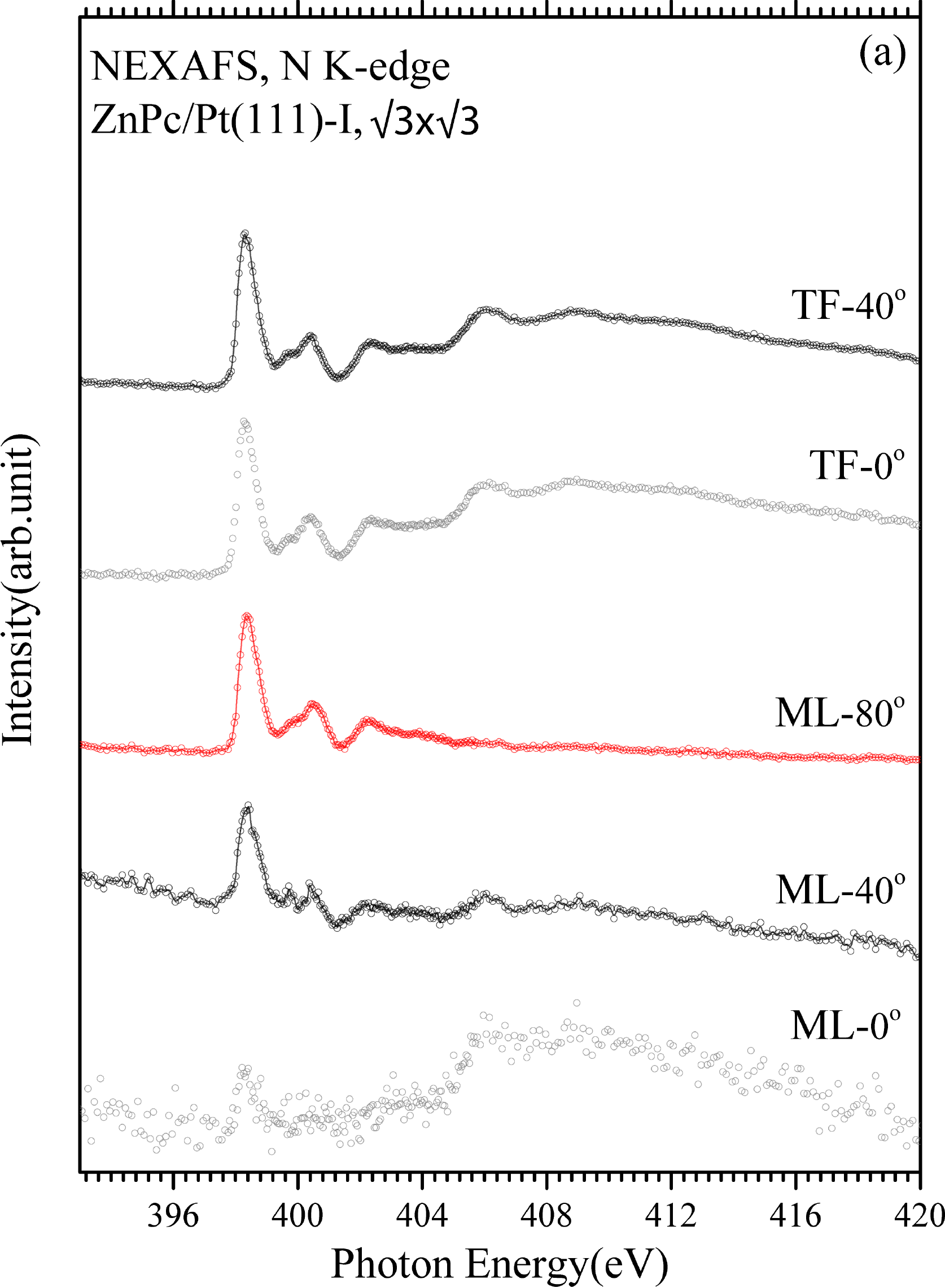}\label{fig:6a}}
\end{minipage}
\hfill
\begin{minipage}[c]{0.493\columnwidth}
\subfloat{\includegraphics[width=\columnwidth]{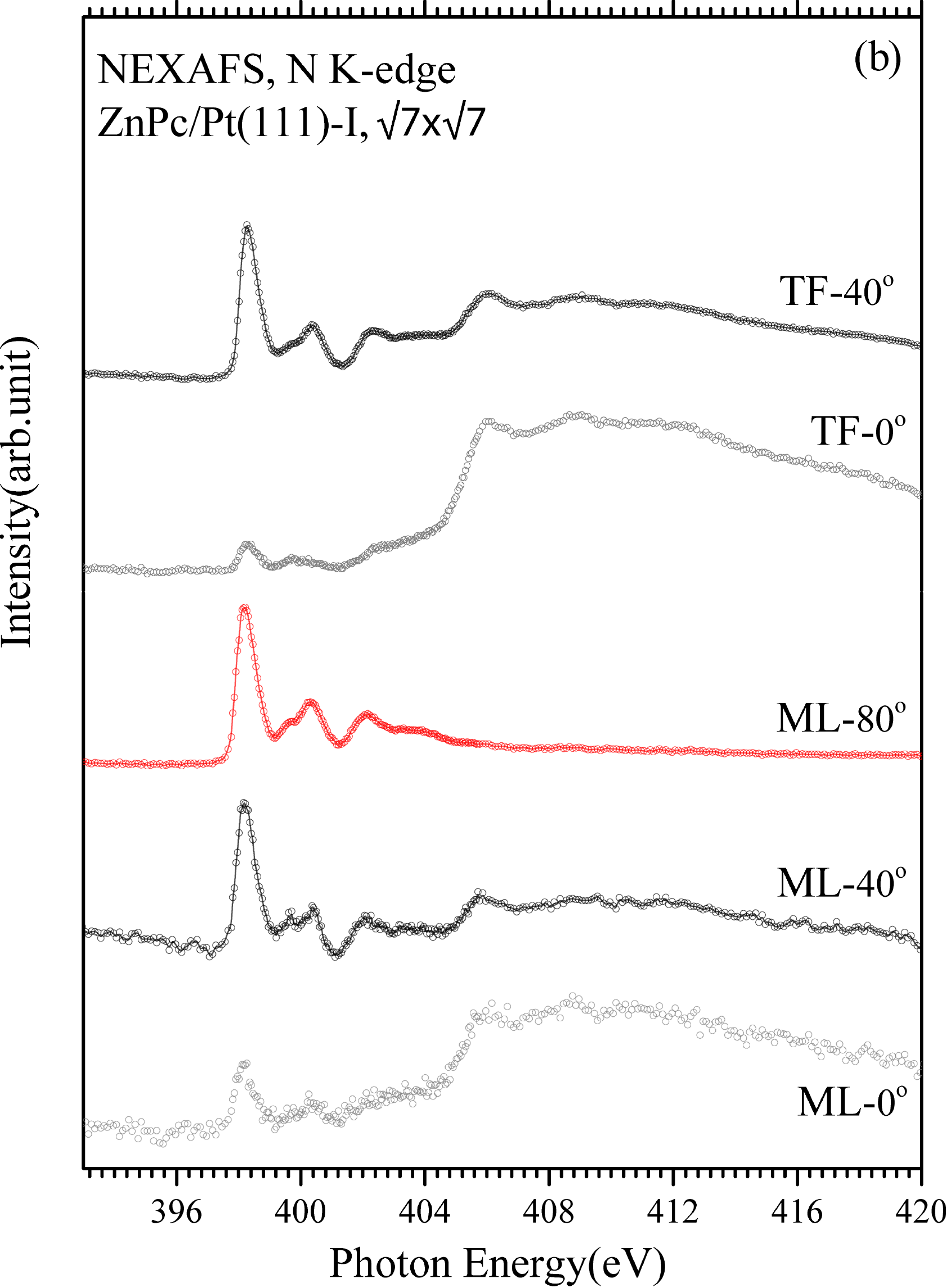}\label{fig:6b}}
\end{minipage}
\caption{N K-edge NEXAFS spectra of monolayers and thick films of ZnPc on (a) 
PtI-$\left(\sqrt{3}\times\sqrt{3}\right)R30^{\circ}$ and (b) PtI-$\left(\sqrt{7}\times\sqrt{7}\right)R19.1^{\circ}$.}
\label{fig:6}
\end{figure}

XAS from ML and TF of ZnPc on Pt-I are presented in Fig.~\ref{fig:6}. They were measured at the
same three angles as for the bare Pt substrate above. On $\left(\sqrt{3}\times\sqrt{3}\right)$ the angular dependence of
the $\pi^{\ast}$ and $\sigma^{\ast}$ intensities implies a flat lying geometry at ML coverage. At increasing
coverage the molecules adopt a slightly tilted geometry. On $\left(\sqrt{7}\times\sqrt{7}\right)$ the monolayer also
appears to lie down flat on the surface and keep this orientation at higher coverage. A closer
look at the first resonance reveals that the extra shoulder on the low-photon-energy side
observed in Fig.~\ref{fig:1} is not present here, indicating that N is less involved in the interfacial
interactions than on Pt(111).

\subsubsection{{\rm Molecular core levels}}

\begin{figure}
\centering
\begin{minipage}[c]{0.498\columnwidth}
\subfloat{\includegraphics[width=\columnwidth]{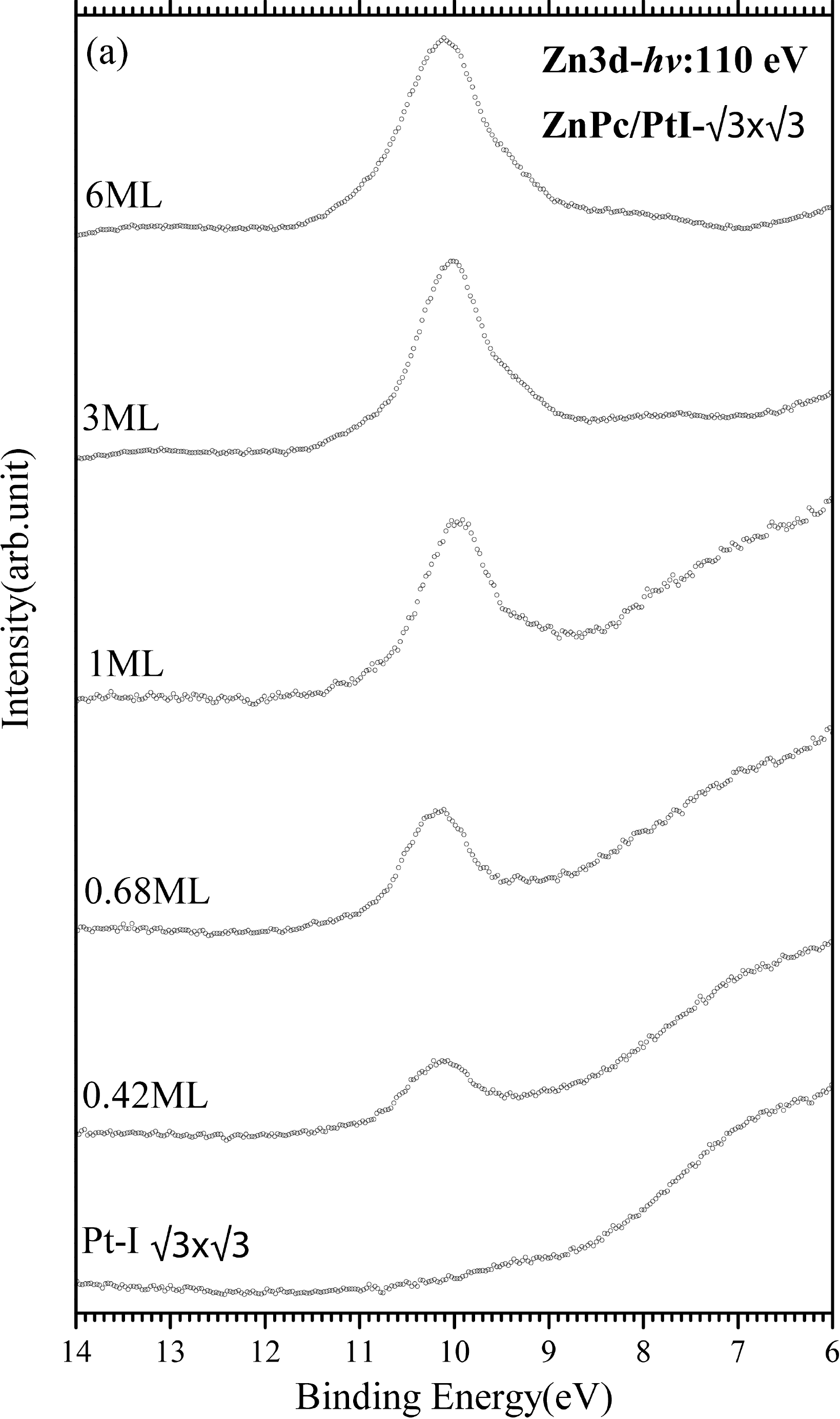}\label{fig:7a}}
\end{minipage}
\hfill
\begin{minipage}[c]{0.488\columnwidth}
\subfloat{\includegraphics[width=\columnwidth]{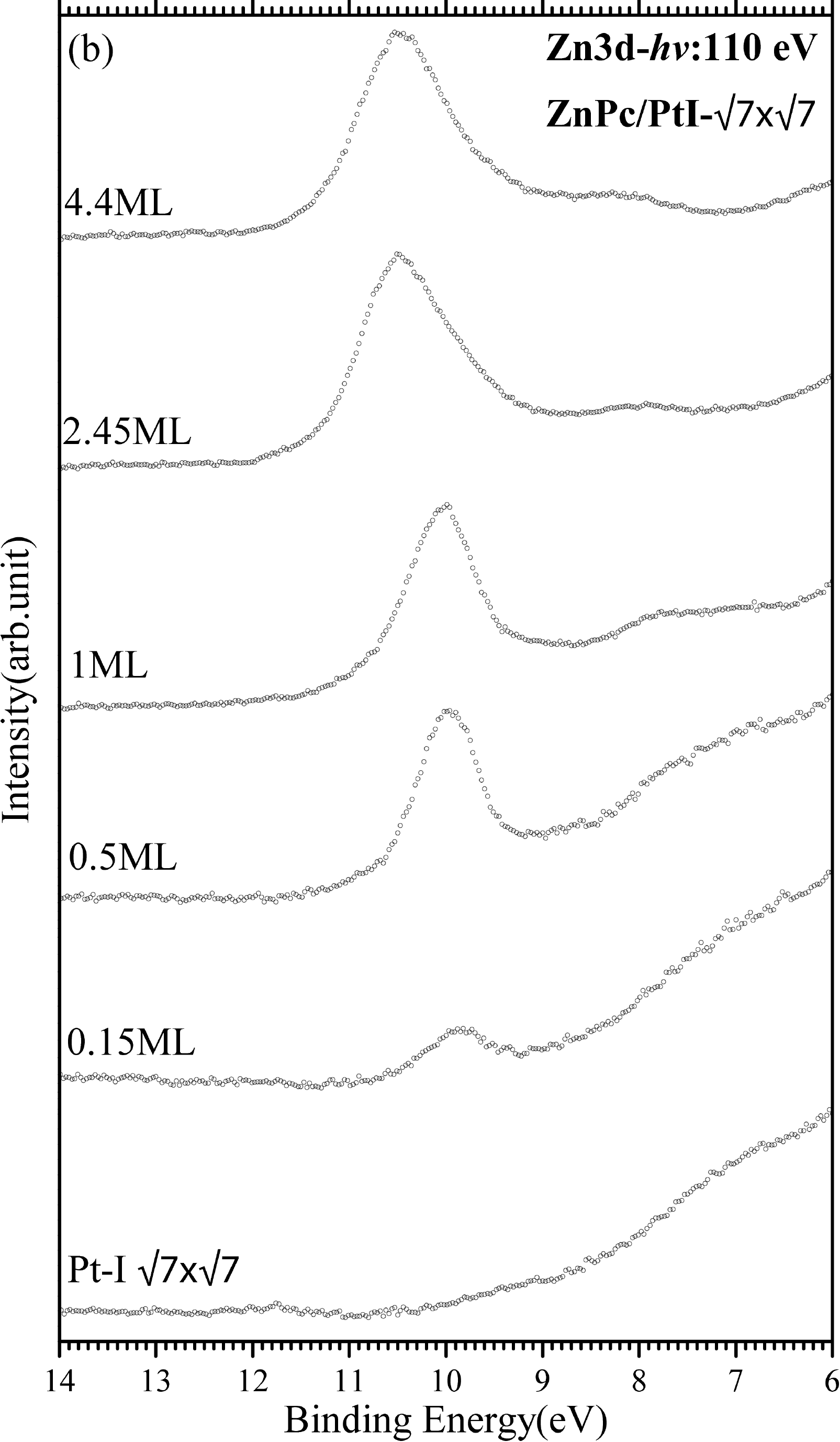}\label{fig:7b}}
\end{minipage}
\caption{Zn$3d$ photoemission spectra of ZnPc layers on (a) PtI-$\left(\sqrt{3}\times\sqrt{3}\right)R30^{\circ}$ and 
(b) PtI-$\left(\sqrt{7}\times\sqrt{7}\right)R19.1^{\circ}$.}
\label{fig:7}
\end{figure}

Figure~\ref{fig:7a} shows Zn$3d$ spectra from ZnPc on $\left(\sqrt{3}\times\sqrt{3}\right)$, where the Zn$3d$ signal 
appears as a single peak at $10.13$~eV. This is $0.65$~eV higher than for ZnPc on Pt(111). In addition, there is
no observable coverage-dependent shift. Figure~\ref{fig:7b} presents the corresponding Zn$3d$ spectra
from ZnPc on $\left(\sqrt{7}\times\sqrt{7}\right)$. The binding energy at the lowest coverage is lower than on
$\left(\sqrt{3}\times\sqrt{3}\right)$, but higher than on bare Pt. Moreover, the thickness-dependent shift is $0.47$~eV
here, larger than on $\left(\sqrt{3}\times\sqrt{3}\right)$ but smaller than on Pt.

\begin{figure}
\centering
\subfloat{\includegraphics[width=0.493\columnwidth]{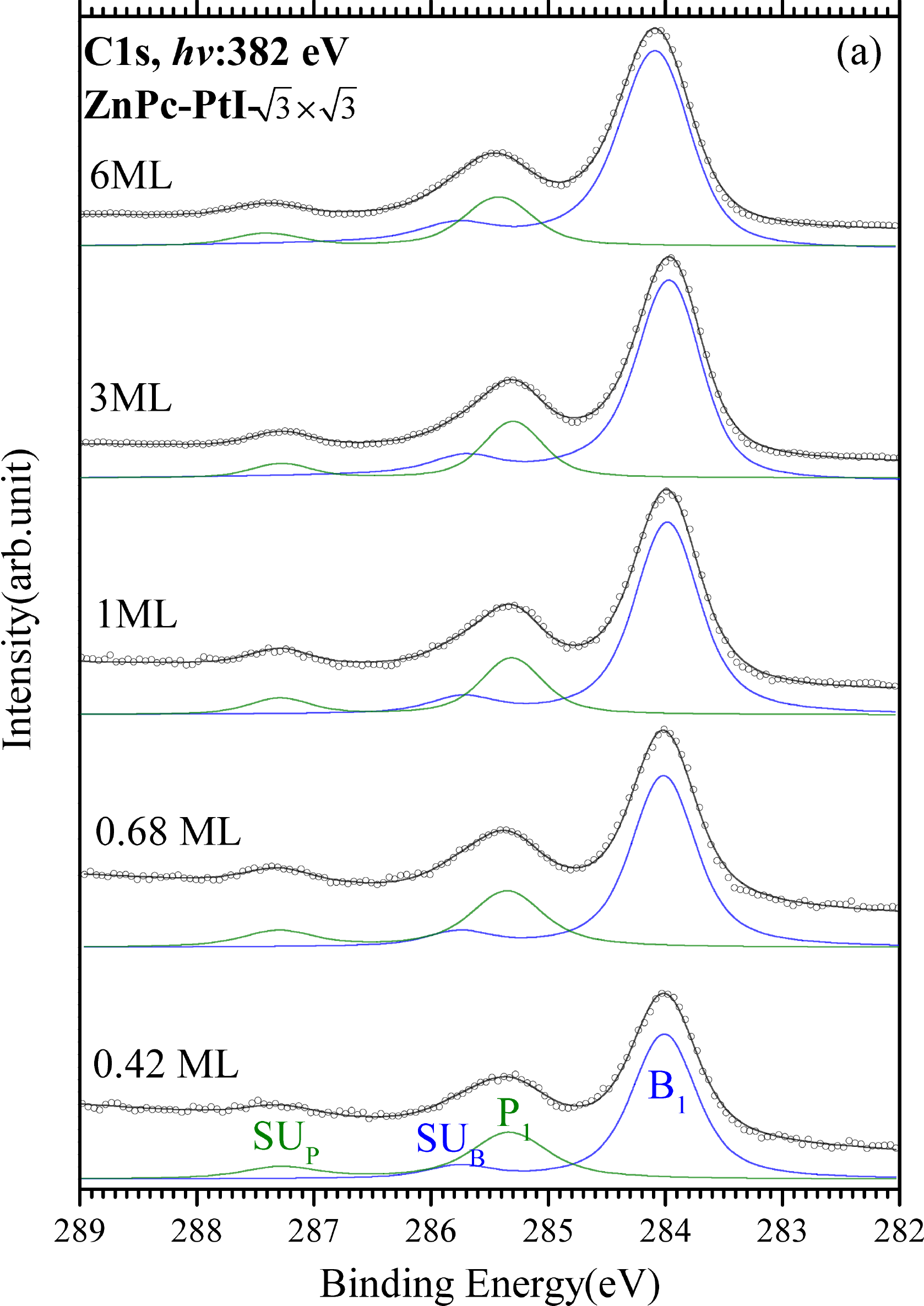}\label{fig:8a}}
\subfloat{\includegraphics[width=0.493\columnwidth]{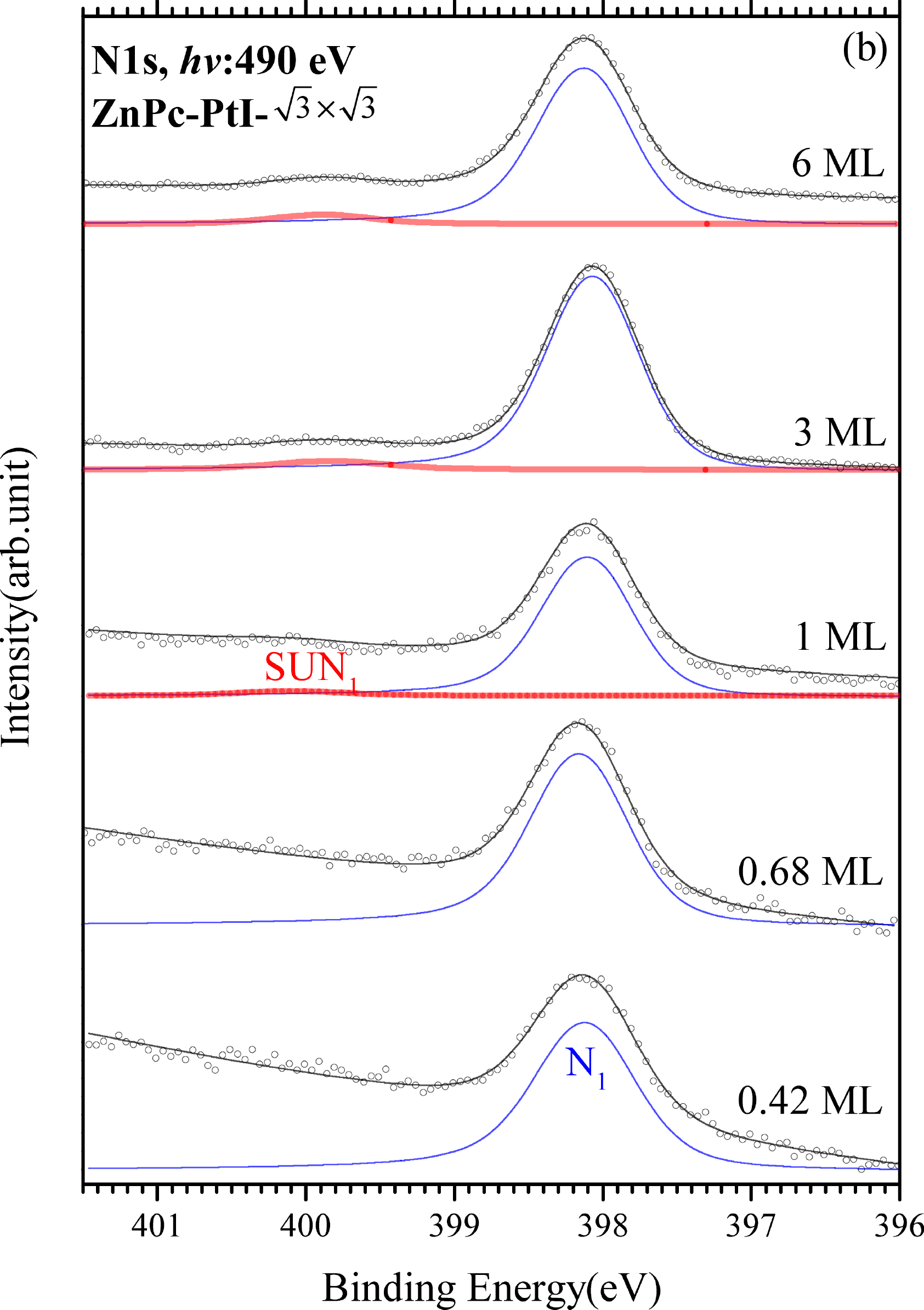}\label{fig:8b}}\\[2mm]
\subfloat{\includegraphics[width=0.493\columnwidth]{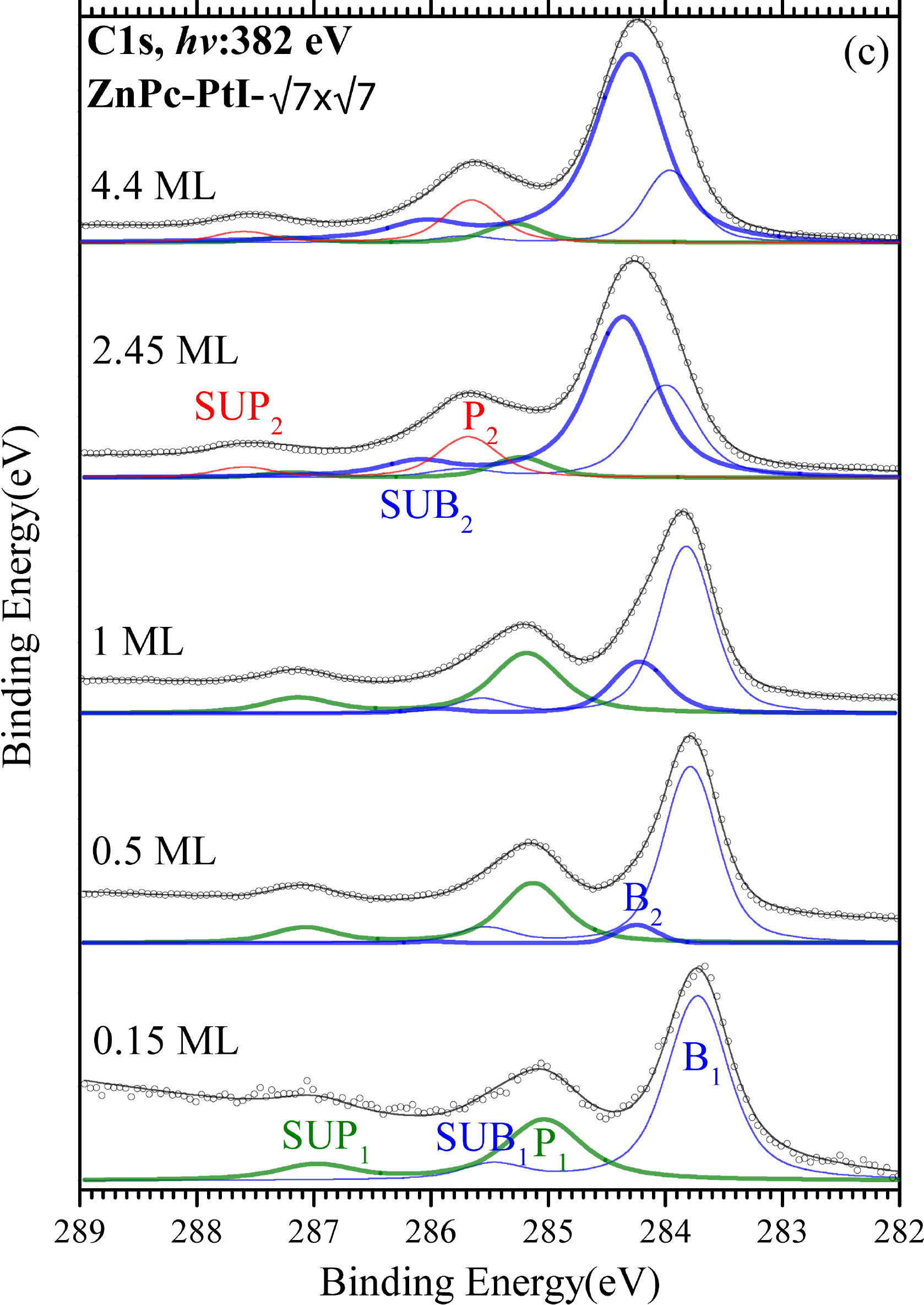}\label{fig:8c}}
\subfloat{\includegraphics[width=0.493\columnwidth]{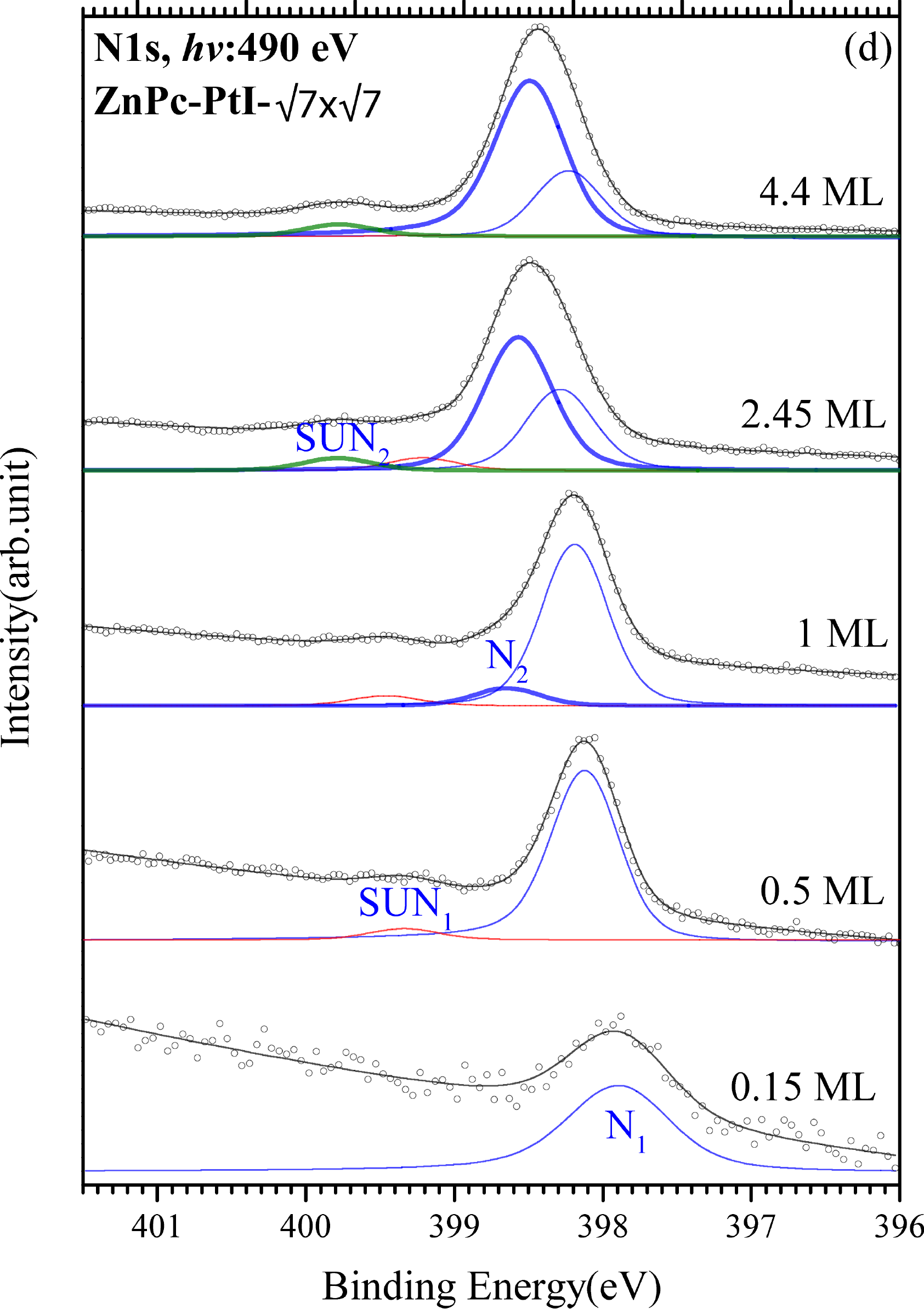}\label{fig:8d}}
\caption{(a) C$1s$ spectra and (b) N$1s$ spectra from ZnPc on PtI-$\left(\sqrt{3}\times\sqrt{3}\right)R30^{\circ}$. (c) 
C$1s$ spectra and (d) N$1s$ spectra from ZnPc on PtI-$\left(\sqrt{7}\times\sqrt{7}\right)R19.1^{\circ}$.}
\label{fig:8}
\end{figure}

C$1s$ measured from increasingly thick ZnPc films on $\left(\sqrt{3}\times\sqrt{3}\right)$ are shown in 
Fig.~\ref{fig:8a}. The same two components (C$_\text{B}$ and C$_\text{P}$) obtained for the thick film on the 
$\left(1\times1\right)$ surface can be fitted to all spectra. There is practically no coverage-dependent shift. 
C$_\text{B}$ and C$_\text{P}$ are located at slightly lower BE than on Pt(111), and they keep their energy separation 
at all coverages. The N$1s$ spectra are presented in Fig.~\ref{fig:8b}. Again, and similar to the C$1s$ and
Zn$3d$ levels, there is no coverage-dependent shift; the binding energy is stable at $398.15$~eV.
The low-coverage binding energies of N$1s$ and C$1s$ are just a little smaller than the low-coverage results from 
$\left(1\times1\right)$, whereas for Zn$3d$ the difference is larger.

C$1s$ and N$1s$ spectra from ZnPc on $\left(\sqrt{7}\times\sqrt{7}\right)$ are presented in Figs.~\ref{fig:8c} 
and~\ref{fig:8d}, respectively. At the lowest coverage, the binding energies are lower than for 
$\left(\sqrt{3}\times\sqrt{3}\right)$. However, with increasing thickness, both C$1s$ and N$1s$ shift to higher binding 
energies. The curve fitting reveals the presence of different components in the first layer, B$_1$, P$_1$ and N$_1$, and
components from the growing film, (B$_2$, P$_2$ and N$_2$). These interface components together
with the coverage-dependent shift suggest a charge transfer from the substrate to the
molecules. The coverage-dependent shifts are $0.52$~eV for benzene carbon, $0.57$~eV for
pyrrole carbon and $0.52$~eV for nitrogen, thus very close to the Zn$3d$ shift of $0.47$~eV.

On Pt(111) a Pt-Zn mediated interaction, including charge transfer and a deformation of
the molecular plane, was suggested, based on different thickness dependent shifts. For both
iodine substrates, the thickness-dependent shifts are about the same for all atoms, in strong
contrast to the findings on Pt(111) (see Table~\ref{tab:II}), indicating a homogeneous charge
distribution and charge transfer screening. Interestingly, for $\left(\sqrt{3}\times\sqrt{3}\right)$ there is no 
coverage dependent shift at all. The origin of the coverage dependent shift is a reduced screening of the
core hole when the distance to the surface increases. Obviously, the charge transport time to
the core hole depends on the conductivity of the organic film, which apparently is higher on
$\left(\sqrt{3}\times\sqrt{3}\right)$ than on $\left(\sqrt{7}\times\sqrt{7}\right)$ than on $\left(1\times1\right)$. The 
conductivity can have different origins; structure within the film~\cite{Yu-JPhysChemC-2011} and doping or perhaps a 
combination.

\subsubsection{{\rm I$4d$}}

\begin{figure}
\centering
\begin{minipage}[c]{0.493\columnwidth}
\subfloat{\includegraphics[width=\columnwidth]{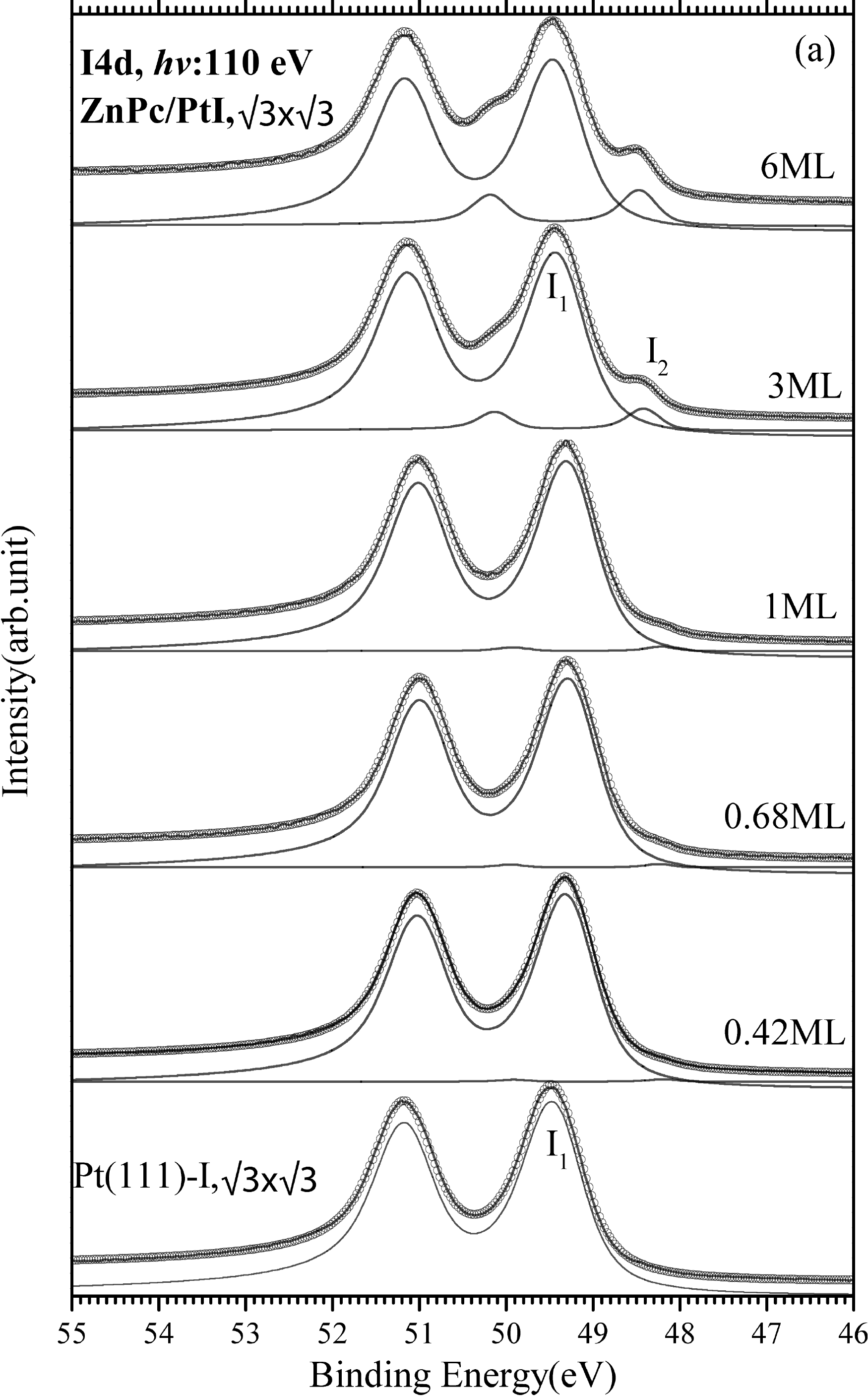}\label{fig:9a}}
\end{minipage}
\hfill
\begin{minipage}[c]{0.493\columnwidth}
\subfloat{\includegraphics[width=\columnwidth]{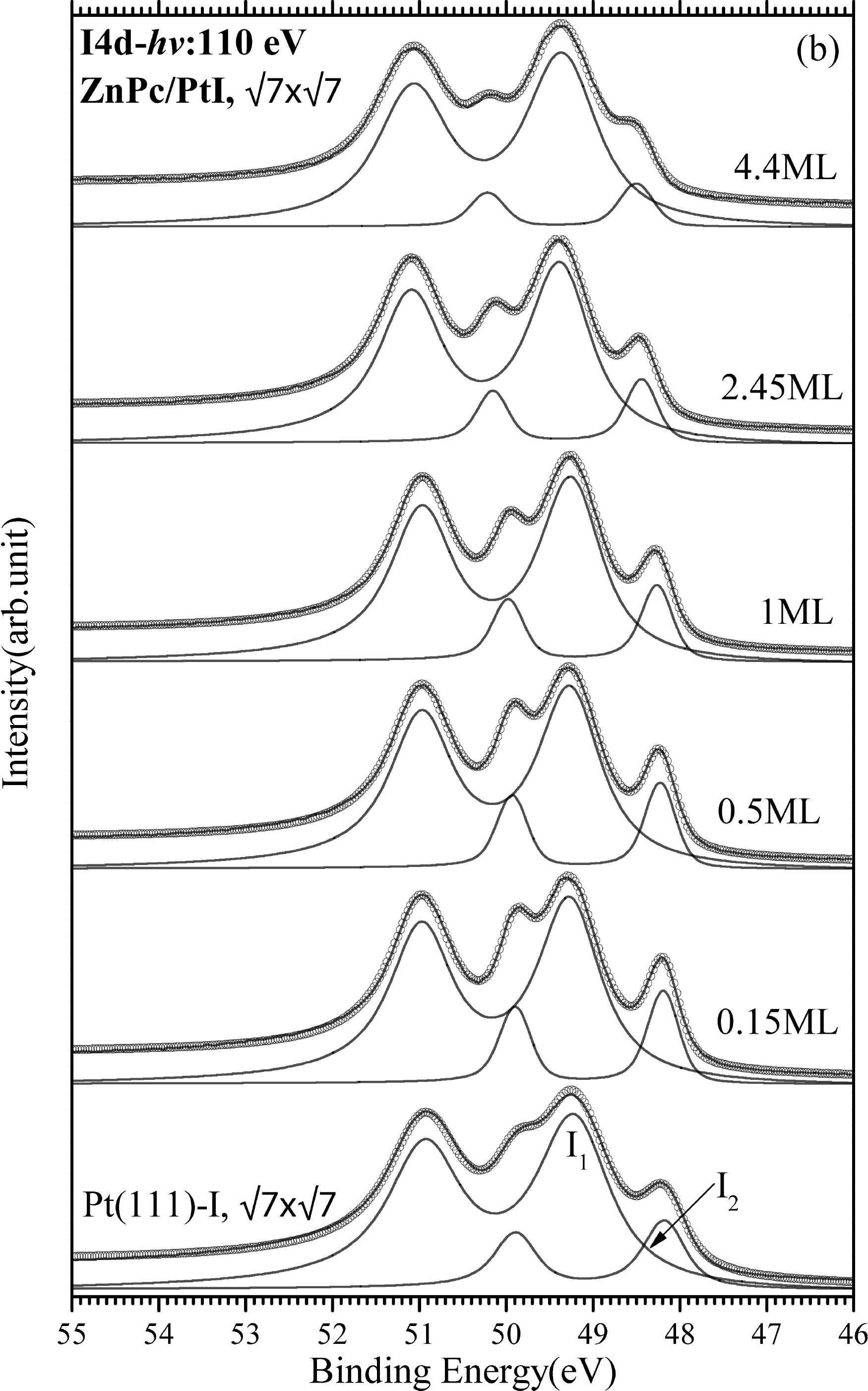}\label{fig:9b}}
\end{minipage}
\caption{I$4d$ spectra from ZnPc on (a) PtI-$\left(\sqrt{3}\times\sqrt{3}\right)R30^{\circ}$ and (b) 
PtI-$\left(\sqrt{7}\times\sqrt{7}\right)R19.1^{\circ}$.}
\label{fig:9}
\end{figure}

The I$4d$ spectra, together with their numerical fits are presented in Fig.~\ref{fig:9}. The 
$\left(\sqrt{3}\times\sqrt{3}\right)$ spectra are presented in the left panel and from 
$\left(\sqrt{7}\times\sqrt{7}\right)$ in the right panel. The amount of
deposited ZnPc is indicated in the figure. The $\left(\sqrt{3}\times\sqrt{3}\right)$ spectrum is fitted with one
spin-orbit doublet, I$_1$ at $51.15$~eV (I$4d_{3/2}$), representing hollow site iodine, while the 
$\left(\sqrt{7}\times\sqrt{7}\right)$ spectrum holds two clearly separate components (I$_1$ and I$_2$) at $50.91$~eV 
and $49.87$~eV binding energies. I$_1$ represents hollow-site iodine and I$_2$ represents top-site iodine. This shift 
has been observed previously~\cite{DiCenzo-PhysRevB-1984,Goethelid-JChemPhys-2012} and it was explained as being 
essentially due to the distance between
the iodine adatom and the surface rather than the ionicity/charge on the adatom~\cite{Bagus-SurfSci-2009}.

The adsorption of ZnPc induces changes in I$4d$ spectra from both surfaces. On $\left(\sqrt{7}\times\sqrt{7}\right)$ 
I$_1$ narrows and shifts slightly to higher binding energy. I$_2$ shifts closer to I$_1$, i.e. even
more to higher binding energy and is also reduced in intensity. On $\left(\sqrt{3}\times\sqrt{3}\right)$ the I$_2$ peak
develops and at the same time both peaks shift gradually to higher binding energy. At the
highest coverage the I$4d$ spectra are rather similar.

The appearance of I$_2$ on the $\left(\sqrt{3}\times\sqrt{3}\right)$ surface may at first seem surprising, since very
small, if any, changes were observed in C$1s$, N$1s$, Zn$3d$ and Pt$4f_{7/2}$ spectra. One further
observation is important: the relative intensity of I$_2$ increases with the ZnPc coverage, thus it
is not an interface effect. Instead, we suggest that iodine is taken from the Pt surface and
dissolved in the ZnPc thin film. Iodine has been used as dopant in organic films for many 
years~\cite{Curry-JChemPhys-1962, Orr-JACS-1979, Schramm-JACS-1980, Waclawek-ThinSolidFilms-1987, 
Nakamura-JpnJApplPhys-1987, Sharma-MatSciEngB-1996, Brinkmann-JPhysChemA-2004, 
Hayashi-JApplPhys-2011,Mizuta-ThinSolidFilms-2012}. Depending on the concentration the conductivity in for example 
PbPc was increased by as much as nine orders of magnitude~\cite{Waclawek-ThinSolidFilms-1987}.

The I$_2$ from the $\left(\sqrt{3}\times\sqrt{3}\right)$ surface component is shifted to lower binding energy, which in 
a simple picture signifies a higher local electron density and/or good conductivity for screening
of the final state core hole. The screening of the core hole cannot be better than when the
iodine is in direct contact with Pt, so we propose that the shift is due to iodine being in a
negatively charged state, i.e. it acts as an acceptor dopant in the ZnPc film. This is in
agreement with previous results in NiPc~\cite{Waclawek-ThinSolidFilms-1987} and 
pentacene~\cite{Brinkmann-JPhysChemA-2004}.

On $\left(\sqrt{7}\times\sqrt{7}\right)$ there are coverage dependent shifts in C$1s$, N$1s$ and Zn$3d$, which suggest
that the iodine doping is lesser in this case. The shifts are smaller than on Pt(111), thus there
is most probably doping. The dopant related I$_2$ component unfortunately overlaps with the
original I$_2$ component, thus iodine doping cannot unambiguously be confirmed by our I$4d$
spectra.

Doping and differences in conductivity will also affect the structure and relative
molecular orientation in the organic film, since the charge distribution within molecules and the
screening of electrostatic forces will be different.

\section{Conclusion}

The electronic structure of ZnPc layers, from sub-monolayers to thick films, on Pt(111) is
studied by means of X-ray photoelectron spectroscopy (XPS), X-ray absorption spectroscopy
(XAS) and scanning tunneling microscopy (STM). Our core-level-spectroscopy results
suggest a non-planar, buckled molecular configuration for ZnPc adsorbed on Pt(111), induced
by the Zn-Pt interaction. It is shown that ZnPc molecules do not form a complete monolayer
on the Pt surface, due to a surface-mediated intermolecular repulsion. STM images confirm
the absence of a well-ordered complete monolayer and depict a porous structure created by
ZnPc molecules on Pt(111). Molecules are lying almost parallel to the substrate at lower
coverage, while they prefer a tilted position at higher coverage, due to the reduced molecule-substrate interaction. Our 
photoemission results illustrate that ZnPc is practically decoupled from Pt, starting from the second layer.
Moreover, monolayers of ZnPc have been adsorbed on two different Pt-I reconstructed surfaces: 
$\left(\sqrt{3}\times\sqrt{3}\right)R30^{\circ}$ and $\left(\sqrt{7}\times\sqrt{7}\right)R19.1^{\circ}$. Iodine, in 
particular the $\left(\sqrt{3}\times\sqrt{3}\right)$ reconstructed surface, reduces the surface-molecule interaction, 
resulting in a more homogeneous charge distribution and a reduction in the molecular distortion. At increasing
ZnPc thickness iodine diffuses from the Pt surface into the molecular film and acts as an
acceptor creating a hole conducting ZnPc film.

\begin{acknowledgments}
The Swedish Energy Agency (STEM), the Swedish Research Council (VR), the G\"oran
Gustafsson Foundation, and the Carl Trygger Foundation are kindly acknowledged for
financial support. The authors would like to also thank kind staff at MAX-lab for technical
support.
\end{acknowledgments}

\end{document}